\documentclass[10pt,conference]{IEEEtran}
\usepackage{cite}
\usepackage{amsmath,amssymb,amsfonts}
\usepackage{algorithmic}
\usepackage{graphicx}
\usepackage{textcomp}
\usepackage{xcolor}
\usepackage{adjustbox}
\usepackage{subcaption}
\begin{document}

\title{Leveraging Commit-Size Context and Hyper Co-Change Graph Centralities for Defect Prediction
}

\author{
  \IEEEauthorblockN{Amit Kumar}
  \IEEEauthorblockA{Dept. of IT, IIIT Allahabad\\
    Prayagraj, India \\
    amitchandramunityagi@gmail.com}
  \and
  \IEEEauthorblockN{Hrishikesh Ethari}
  \IEEEauthorblockA{Dept. of CSE, IIIT Manipur\\
    Imphal, India \\
    hrishikeshethari@gmail.com}
  \and
  \IEEEauthorblockN{Sonali Agarwal}
  \IEEEauthorblockA{Dept. of IT, IIIT Allahabad\\
    Prayagraj, India \\
    sonali@iiita.ac.in}
}

\maketitle
\begin{abstract}
File-level defect prediction models traditionally rely on product and process metrics. While process metrics effectively complement product metrics, they often overlook \textit{commit size}—the number of files changed per commit—despite its strong association with software quality. Network centrality measures on dependency graphs have also proven to be valuable product-level indicators. Motivated by this, we first redefine process metrics as \textit{commit-size–aware process metric vectors}, transforming conventional scalar measures into 100-dimensional profiles that capture the distribution of changes across commit-size strata. We then model change history as a \textit{hyper co-change graph}, where hyperedges naturally encode commit-size semantics. Vector centralities computed on these hypergraphs quantify size-aware node importance for source files. Experiments on nine long-lived Apache projects (32 releases) using five popular classifiers show that replacing scalar process metrics with the proposed commit-size–aware vectors, alongside product metrics, consistently improves predictive performance. Median gains range from 3.5--7.3\% in AUROC, 5.6--26.7\% in AUPRC, 8.0--17.6\% in F1, and 11.5--28.3\% in MCC, while Brier scores decrease by 3.0--11.3\%. Incorporating hyper co-change graph centralities with product and vector process metrics achieves the best overall results—yielding the highest AUROC, MCC, and F1 in 75.6\%, 66.7\%, and 71.1\% of cases, respectively. Friedman and post-hoc Nemenyi tests confirm these improvements are statistically significant ($p < 0.05$). These findings establish that commit-size–aware process metrics and hypergraph-based vector centralities capture higher-order change semantics, leading to more discriminative, better-calibrated, and statistically superior defect prediction models.

\end{abstract}

\begin{IEEEkeywords}
Software defect prediction, process metrics, commit-size–aware process metrics, hyper co-change graph, vector centrality.
\end{IEEEkeywords}

\section{Introduction}
\par Software Defect prediction models primarily use two types of metrics: product and process metrics. Product metrics, derived from code analysis, characterize the structural properties of software using measures such as McCabe’s cyclomatic complexity \cite{mccabe1976complexity} and the CK suite of object-oriented metrics \cite{chidamber1994metrics}, among numerous metrics proposed to evaluate software quality. In contrast, process metrics, such as commit count and change entropy \cite{hassan2009predicting,rahman2013and}, are derived from the project’s historical change data and capture the temporal dynamics of software evolution, reflecting the intensity, scope, and distribution of code modifications. Numerous studies have shown that process metrics significantly enhance the performance of defect prediction and effectively complement product metrics \cite{rahman2013and,bird2009does,majumder2022revisiting}. Moreover, they are language-independent and adaptable across diverse projects, making them an indispensable component of defect prediction models.
\par
Although conventional process metrics for file-level defect classification capture important aspects of software change behavior—such as the number of commits and the number of developers who modified a file—they generally overlook an important factor: the size of the commits (i.e., the number of source code files changed in each commit) that affected the file, which is closely associated with software quality. Prior research has demonstrated that commit size, measured by the number of files changed, influences design quality \cite{hammad2011automatically}, affects the success of automated software repair \cite{le2013current}, and is correlated with bug reopens, as bugs fixed in larger commits tend to be reopened more often \cite{shihab2013studying}. Larger commits have also been identified as indicators of code decay, architectural degradation, and erosion of software design quality \cite{eick2002does,bandi2013empirical,li2011case}, and are more bug-inducing than smaller ones \cite{sliwerski2005changes}.
\par
However, while the number of files changed in a commit (NF) has been a widely used metric in \textit{Just-In-Time (JIT) defect prediction}---which predicts whether a commit made or proposed by a developer is defect-prone \cite{ni2022just,kamei2012large,khanan2020jitbot,fan2019impact}---it is rarely incorporated into file-level defect prediction. Existing process metrics typically rely on coarse-grained measures, such as the number of commits or the total lines added and deleted in the previous release. However, commits of different sizes can introduce defects with varying likelihoods and impact \cite{misirli2016studying}, and aggregating such change information into simple count-based metrics obscures these differences.
\par
\begin{figure}[htbp]
    \centering
    \begin{adjustbox}{max width=\linewidth}
        \includegraphics{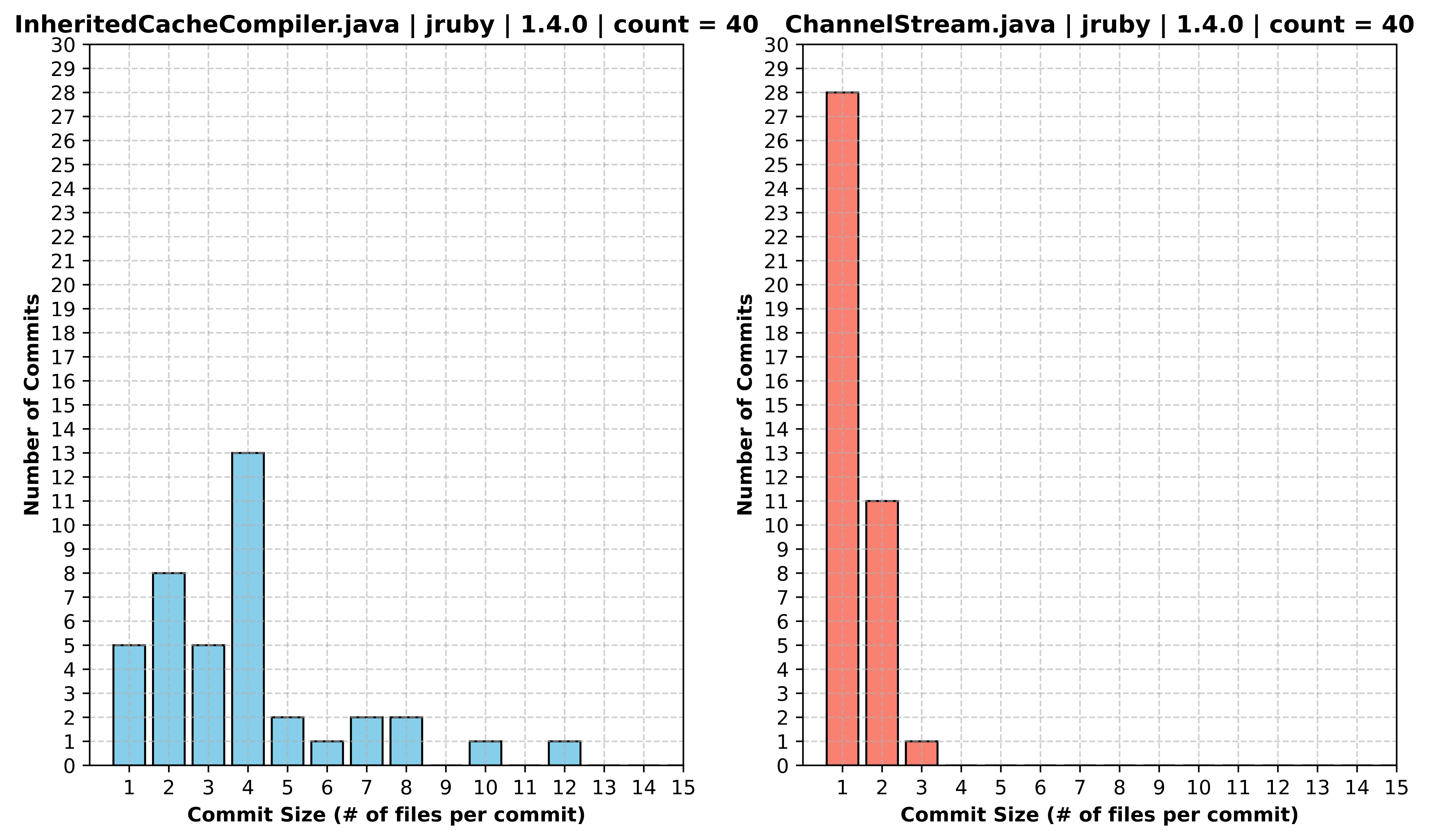}
    \end{adjustbox}
    \caption{Commit size distribution for two files with identical commit counts in JRuby~1.4.0.}
    \label{fig:commitsizedistribution}
\end{figure}

Figure~\ref{fig:commitsizedistribution}\footnote{Refer to the digital version for better figure clarity.} illustrates this with an example where both \textit{InheritedCacheCompiler.java} and \textit{ChannelStream.java} in release~1.4.0 of the JRuby project (analyzed in this study) were modified 40~times, yielding identical commit counts. However, \textit{ChannelStream.java} was changed predominantly in smaller commits, whereas \textit{InheritedCacheCompiler.java} was affected by several large commits. A commit-size–agnostic metric such as commit count treats these files as equivalent, overlooking the underlying distribution of commit sizes that may influence their differing defect proneness.

\par
This observation motivates the incorporation of commit-size information into process metrics to improve file-level defect prediction. To this end, we extend conventional process metrics to be \textit{commit-size–stratified}, thereby enhancing their explanatory and predictive capabilities. Specifically, we transform widely used scalar process metrics \cite{rahman2013and,majumder2022revisiting} into \textit{commit-size–stratified change profiles} represented as vectors. For each file, every component of the vector corresponds to the value of the metric computed exclusively from commits within a specific commit-size bin. This stratification preserves the distributional characteristics of the metric across commits of different magnitudes—contextual information that a single scalar aggregation inherently discards.
\par
The contribution of this study is an empirical evaluation of whether these commit-size–aware vector process metrics, when integrated with product metrics, can enhance the discriminatory power and predictive performance of file-level defect prediction models compared to their scalar counterparts. Accordingly, we formulate our first research question as follows:
\begin{center}
\fbox{\begin{minipage}{\dimexpr\columnwidth-2\fboxsep-2\fboxrule}
\textbf{RQ1:} \textit{Do commit size–aware vector process metrics complement product metrics more effectively than their scalar counterparts in improving defect prediction performance?}
\end{minipage}}
\end{center}
\par
Software entities that frequently change together are said to exhibit a \textit{change dependency} \cite{zimmermann2005mining}. The impact of such co-change relationships on software quality and defect occurrence has been extensively studied. These relationships, where files are modified together within the same commit, are typically modeled as \textit{co-change graphs}. Prior studies have shown that analyzing the structural and topological properties of these graphs aids in understanding software evolution and predicting future defects \cite{d2009relationship,silva2019co,hrishikesh2025co,kouroshfar2013studying,kirbas2017relationship}.
\par
Some of the most popular measures or metrics in graph theory come from Social network anlaysis (SNA) which are node centrality measures like degree centrality, betweenness centrality, etc., to measure the node importance in the network. They measure the importance of the node in the network. The node centrality measures on structural/code dependency networks have been used extensively as metrics for defect prediction in the source code files \cite{zimmermann2008predicting,nguyen2010studying,amasaki2020cross,gong2021revisiting}. However, although effective on structural dependency networks, the utility of these centrality measures for defect prediction on co-change networks remains under-evaluated.
\par Simple co-change (pairwise) networks ignore commit size, so they cannot distinguish a file co-changed in numerous small commits from one co-changed in a few large commits. In a hypernetwork\cite{lee2025survey}\footnote{“Hypernetworks” and “hypergraphs” are used synonymously in this paper}, the number of nodes participating in an edge is not limited to two, extending the pairwise model. Representing each commit as a hyperedge linking all modified files captures higher-order group interactions and inherently encodes commit-size semantics.
\par
Among the various node importance measures proposed for hypernetworks, \textit{vector centrality} \cite{kovalenko2022vector} is particularly well suited for modeling co-change behavior and capturing commit-size semantics. Unlike traditional scalar measures, it represents each node’s importance as a vector across hyperedges of different sizes, reflecting its diverse roles in development and maintenance. Higher values in lower dimensions indicate frequent participation in small commits involving few files, whereas higher-dimensional values signify involvement in large, system-wide, or \textit{cross-cutting} modifications that span multiple modules. Consequently, vector centrality captures a file’s participation across diverse commit contexts and, in doing so, naturally integrates commit size semantics for a richer change-history profile.
\par Motivated by the ability of hypernetworks to capture higher-order interactions and the potential of vector centrality to integrate commit-size semantics into node importance, we pose our second research question as follows:\\\\
\fbox{\begin{minipage}{\dimexpr\columnwidth-2\fboxsep-2\fboxrule}
\textbf{RQ2:} \textit{What impact does incorporating vector centralities derived from hyper-co-change networks have on the performance of defect prediction models built upon product and vector process metrics?}
\end{minipage}} \\
\par
In summary, this paper makes the following key contributions:
\begin{itemize}
\item \textbf{Commit size–aware vectorization of process metrics:}
To the best of our knowledge, we are the first to incorporate the \textit{size of the commit} into the computation of process metrics by redefining traditional scalar metrics as \textit{commit size–aware vectors}. This formulation captures file-level change behavior across varying commit magnitudes, preserving richer process semantics. Empirical evaluation across nine long-lived projects (32 releases) shows that the proposed vector process metrics complement product metrics more effectively than their scalar counterparts, leading to significant improvements in defect prediction.
\item \textbf{Hyper–co-change network centralities for enhanced prediction:}
To the best of our knowledge, we are the first to model co-change activities as a \textit{hypergraph} and employ \textit{vector centralities} (degree, betweenness, closeness, eigenvector) to compute commit size–aware node importance measures for defect prediction. These hypergraph-based vector centralities capture higher-order co-change relationships among files modified together in commits. Incorporating them alongside product and vector process metrics yields the best overall prediction performance, with statistically significant gains validated through Friedman and post-hoc Nemenyi analyses.
\end{itemize}


\begin{figure*}[htbp]
    \centering
    \begin{adjustbox}{max width=\linewidth}
    \includegraphics{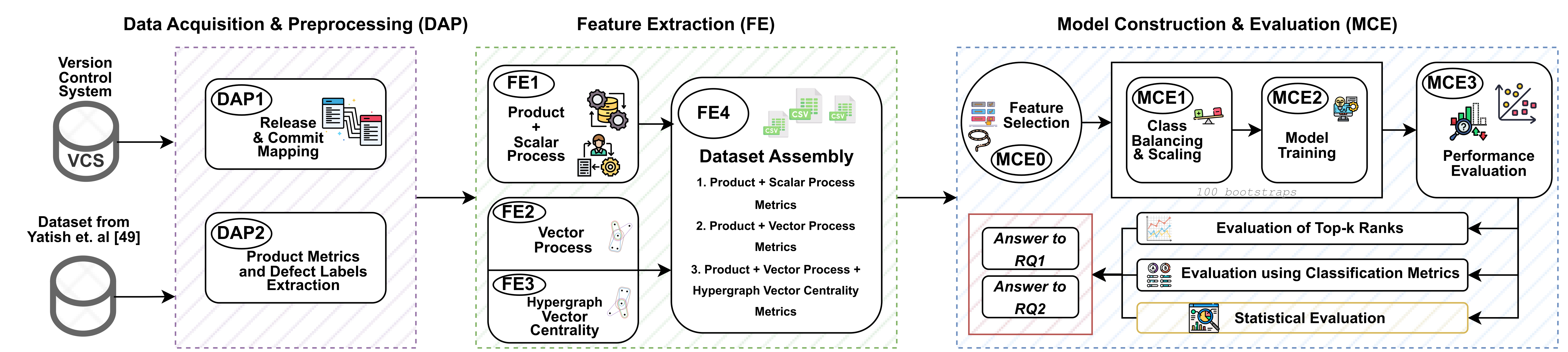}
    \end{adjustbox}
    \caption{Schematic representation of the overall methodological framework}
    \label{fig:schematic}
\end{figure*}
\section{Related Work}
There are two prominent categories of metrics used in software defect prediction: \textit{product} and \textit{process} metrics. Product metrics, derived from static code analysis, characterize the structural properties of software, whereas process metrics, such as commit count, are obtained from a project’s historical change data. Prior research has shown that process metrics often outperform product metrics in predictive accuracy. Rahman and Devanbu~\cite{rahman2013and} provided early evidence of this trend, later reinforced by large-scale studies such as Majumder et al.~\cite{majumder2022revisiting}. Moreover, Majumder et al.~\cite{majumder2024less} demonstrated that process metrics remain effective even under semi-supervised settings with limited labeled data.
\par
Changes made to the system form the foundation of process metrics. Hassan~\cite{hassan2009predicting} showed that systems with more distributed changes across files are more defect-prone, suggesting that file-level change information is a valuable defect indicator. Nagappan et al.~\cite{nagappan2010change} found that change bursts—frequent changes within short time intervals—are strong predictors of defects. Similarly, Wen et al.~\cite{wen2018well} employed recurrent neural networks to model sequential change patterns and demonstrated that both the amount and the temporal sequence of changes influence defect likelihood.
\par
Several studies have also highlighted that the number of files co-changed with a given file (i.e., commit size) is indicative of post-release defects and software quality. Moser et al.~\cite{moser2008comparative} incorporated the average and maximum number of co-changed files as features and found them useful for defect prediction. Kouroshfar et al.~\cite{kouroshfar2015study} observed that the number of co-changed files can serve as an indicator of architectural quality. Shihab et al.~\cite{shihab2011high} noted that high-impact bugs tend to involve larger commits, while Yan et al.~\cite{yan2017file} proposed using the number of files modified to estimate defect-fixing effort in effort-aware defect prediction. However, to the best of our knowledge, no prior work has comprehensively modeled the \textit{distribution} of commit sizes a file has experienced or utilized its commit-size-aware change history for defect prediction. Our study fills this gap by constructing a commit-size–aware change profile for each file and leveraging it for defect prediction. 
\par
More closely related to our work is by \v{S}iki{\'c} et al.~\cite{vsikic2021improving} who considered aggregated change metrics based on the sizes of commits involving each file. However, they summarized commit sizes using average values, overlooking the fine-grained distribution of small versus large commits, which can be crucial for understanding a file’s defect-proneness. In contrast, our approach preserves the more detailed distributional information, enabling a richer representation of a file’s evolutionary behavior.
\par
Commit size has also been extensively studied in \textit{Just-In-Time (JIT) defect prediction} ~\cite{ni2022just,kamei2012large,khanan2020jitbot,fan2019impact}. However, since our focus is on file-level defect prediction, we only briefly refer to this line of work in this section without discussing it in detail.
\par
Applying network analysis and centrality measures for defect prediction at the product level or within developer communication networks is not new. Network centrality measures on code dependency graphs and developer networks have been extensively applied in prior research~\cite{zimmermann2008predicting,meneely2008predicting,pinzger2008can,nguyen2010studying,gong2021revisiting}. However, since our work focuses primarily on process metrics, we do not discuss these studies in detail.
\par
Simple co-change networks and analyses on them have been used in prior studies to investigate software quality~\cite{silva2015co,silva2019co,silva2014assessing,kouroshfar2013studying}. However, simple pairwise networks cannot accommodate commit size, and this aspect is therefore lost. Our work extends this research direction by leveraging \textit{hyper co-change networks} that capture higher-order relationships among files modified together in a commit. Using this hypergraph representation, we compute \textit{vector centrality}~\cite{kovalenko2022vector} for each file, capturing its role across commits of different sizes and thus encoding both structural and semantic aspects of changes. 
\par 
Finally, hypergraph-based models have recently gained attention in software engineering~\cite{cui2024three,wang2024hecs,rong2022modeling,ersoy2016using,jiang2019inferring}. For instance, Qiu et al.~\cite{qiu2024code} used a hypergraph neural network for defect prediction, though their model mainly focuses on code-level (product) features rather than process data. In contrast, we model co-changes as a hypergraph and employ vector centrality to measure node importance. To the best of our knowledge, this is the first study to apply hypergraph-based vector centrality on process data for defect prediction.
\par
In summary, our research advances prior work by incorporating commit context—specifically, the size of commits in which files were changed—through vector process metrics and by modeling higher-order co-change relationships using a hyper-co-change graph for defect prediction.
\section{Methodology}
In this section, we outline the methodology used to address our research questions. We first describe the dataset and preprocessing steps, followed by the classification models and validation techniques. Finally, we detail the experimental setup employed to answer  our research questions. The overall workflow of the study is illustrated in Figure~\ref{fig:schematic}, which provides a schematic overview of the complete methodological framework.

\subsection{Dataset Description and Preprocessing}
We used the dataset prepared by Yatish et al. \cite{yatish2019mining} for our experiments. This dataset was chosen because the authors carefully selected projects and releases where most issue reports are either closed or fixed and linked to their corresponding code changes, ensuring high data quality. Additionally, it provides 54 product metrics essential for our analysis. Since our objective is to examine whether incorporating commit-size information into process metrics and size-aware node centrality measures complements existing product metrics to enhance defect prediction performance, this dataset is ideally suited for our study.  It is also widely used in prior defect prediction research. \cite{wattanakriengkrai2020predicting,lee2023empirical,gong2021revisiting,jiarpakdee2020empirical,li2022robust,moussa2022use,thongtanunam2024code}. The details of our dataset is shown in table \ref{tab:dataset table}. We primarily used this dataset to obtain product metrics for source code files and their post-release defect labels. However, all process metrics were computed separately from the commit data after cloning the projects.
\begin{table*}[htbp]
\caption{Dataset Description}
\label{tab:dataset table}
\begin{tabular}{ccccccc}
\hline
\textbf{Project} & \textbf{Description} & \textbf{Commits} & \textbf{Src Files} & \textbf{Bugs} & \textbf{Defective Rate} & \textbf{Releases} \\ \hline
\textbf{Activemq} & Messaging and Integration Patterns server & 174-634 & 88-1847 & 1626 & 15.54\%-67.05\% & 5.0.0, 5.1.0, 5.2.0, 5.3.0, 5.8.0 \\
\textbf{Camel} & Enterprise Integration Framework & 420-4937 & 516-1935 & 1271 & 6.4\%-37.79\% & 1.4.0, 2.9.0, 2.10.0, 2.11.0 \\
\textbf{Derby} & Relational Database & 640-1155 & 207-1933 & 4362 & 33.37\%-52.47\% & 10.2.1.6, 10.3.1.4, 10.5.1.1 \\
\textbf{Groovy} & Java-syntax-compatible OOP for JAVA & 137-345 & 93-245 & 382 & 18.28\%-29.73\% & 1.5.7, 1.6.0.Beta1, 1.6.0.Beta 2 \\
\textbf{HBase} & Distributed Scalable Data Store & 536-942 & 540-900 & 2632 & 35.74\%-38.67\% & 0.94.0, 0.95.0, 0.95.2 \\
\textbf{Hive} & Data Warehouse System for Hadoop & 245-524 & 266-1113 & 767 & 15.27\%-41.73\% & 0.9.0, 0.10.0, 0.12.0 \\
\textbf{JRuby} & Ruby Programming Lang for JVM & 687-2674 & 318-501 & 637 & 13.77\%-36.16\% & 1.1, 1.4, 1.5, 1.7 \\
\textbf{Lucene} & Text Search Engine Library & 195-792 & 299-1080 & 1387 & 6.57\%-59.2\% & 2.3.0, 2.9.0, 3.0.0, 3.1.0 \\
\textbf{Wicket} & Web Application Framework & 1-1370 & 735-782 & 479 & 8.3\%-13.14\% & 1.3.0.beta1, 1.3.0.beta2, 1.5.3 \\ \hline
\end{tabular}
\end{table*}
\par To compute the conventional and vector process metrics, as well as the hypergraph-based vector centralities of source code files derived from co-change information, we first analyze the version control data—specifically, the commits and their associated metadata. However, in defect prediction studies, a recurring challenge lies in deciding which commits should be included or excluded when computing features and constructing the training dataset for predicting future defects. Selecting an appropriate threshold is critical: very large commits are often associated with administrative activities such as repository restructuring or dependency refactoring \cite{mahbub2023defectors}. Including such commits can inflate noise and lead to false positives by incorrectly labeling defect-free files as defective, whereas using too restrictive a threshold may exclude relevant commits and overlook genuinely defective files \cite{zhou2025bridging}. Consequently, several recent studies have adopted a standard threshold of 100 files per commit, excluding commits that modify more than 100 files from defect prediction datasets \cite{mcintosh2018fix,keshavarz2022apachejit,ni2022best,pan2023fine,mahbub2023defectors}. Following this convention, we also excluded all commits that changed more than 100 files from our final dataset. For files that have not been modified in any commit of size 100 or larger, we applied a zero-padding strategy to make the vector centrality vectors uniform.
\subsection{Classification and Model Validation}
\par We employ five widely used classifiers in our study—Logistic Regression, Support Vector Machine (SVM), XGBoost, Gradient Boosting Machine (GBM), and Random Forest. Logistic Regression serves as a strong, interpretable linear baseline with built-in regularization and well-calibrated probabilities. SVM is a margin-based learner that performs effectively in high-dimensional, sparse feature spaces common to software metrics. XGBoost, a regularized gradient-boosted tree model, captures complex feature interactions and handles missing values efficiently. GBM offers a flexible boosting framework that balances bias and variance, yielding robust and calibrated predictions. Random Forest, an ensemble of bagged decision trees, mitigates overfitting and provides stable feature importance estimates under noisy metric sets. Collectively, these models span linear and ensemble-based paradigms and have been extensively validated in prior defect prediction studies \cite{gong2022comprehensive,shin2021explainable,rajbahadur2017impact,wang2022machine}. We used the \texttt{Scikit-learn} library \cite{pedregosa2011scikit} implementation for all the classifiers in our study.
\par To ensure robustness and minimize bias in our results, we employ an out-of-sample bootstrap validation technique. This method has been widely recommended for producing stable and unbiased estimates in defect prediction studies \cite{yatish2019mining,gong2021revisiting,tantithamthavorn2016empirical,jiarpakdee2019impact,gong2022comprehensive}. For a dataset with $D$ instances, each bootstrap sample is generated by randomly selecting $D$ instances with replacement, leaving out roughly 36\% of the instances as unseen (out-of-bag) test data. The model is trained and evaluated on 100 such bootstrap samples, and the final performance is reported as the average across these iterations.
\subsection{Handling Data Imbalance and Feature Selection}
Before training classifiers and evaluating their performance, we perform two key preprocessing steps. First, since defect datasets often suffer from class imbalance—where buggy files are significantly fewer than non-buggy ones—we apply the Synthetic Minority Over-sampling Technique (SMOTE) \cite{chawla2002smote} to balance the data. SMOTE is widely and successfully used in defect prediction studies \cite{majumder2022revisiting,wang2013using}.
\begin{sloppypar}
To reduce noise and redundancy among features—which are often interdependent—we apply feature selection before training the classifiers. Several feature selection techniques have been explored for defect prediction \cite{xu2016impact}. Xu et al. \cite{xu2016impact} reported that Correlation-based Feature Selection (CFS) performs well in many cases, though its effectiveness varies across datasets. Jiarpakdee et al. \cite{jiarpakdee2018autospearman} proposed \textit{AutoSpearman} as an effective feature selection method for defect prediction. However, since correlation-based feature selection methods and AutoSpearman rely on pairwise correlations, they do not scale efficiently to high-dimensional settings such as ours, which involves over 1,800 features. Therefore, we adopt the Hilbert–Schmidt Independence Criterion (HSIC) Lasso \cite{yamada2014high}, a kernel-based method well suited for high-dimensional data that can capture nonlinear dependencies between features and target variables and is widely used for feature selection in machine learning. 
\end{sloppypar}
\subsection{Experimental Setup to Answer RQ1}
\label{RQ2_setup}
\subsubsection{Details of the Metrics}
\begin{table*}[t]
\centering
\scriptsize
\caption{Vector Process Metrics}
\label{tab:vectorprocess4}
\begin{tabular}{|p{2.5cm}|p{4.7cm}|p{2.9cm}|p{6cm}|}
\hline
\textbf{Metric} & \textbf{Definition} & \textbf{Formula used to compute $i^{th}$ component for the vector} & \textbf{Notations} \\
\hline

\textbf{Commit Count Vector} ($\mathbf{V}^{(comm)}_f$) &
Distribution of how often file $f$ appears in commits of different sizes. &
\parbox[c]{2.9cm}{\centering 
\(\mathbf{V}^{(\text{comm})}_f[i] = \dfrac{|C_{f,i}|}{|C_f|}\)} &
\parbox[t]{6cm}{$C_{f,i}$ = commits of size $i$ involving  file $f$; \\$C_f$ = total commits involving  $f$.} \\

\hline

\textbf{Developer Activity Vector} ($\mathbf{V}^{(adev)}_f$) &
Distribution of the number of developers who modified file $f$ across commits of different sizes in the target release. &
\parbox[c]{2.9cm}{\centering 
\(\mathbf{V}^{(\text{adev})}_f[i] = \dfrac{|D_{f,i}|}{|D_f|}\)} &
\parbox[t]{6cm}{$D_{f,i}$ = developers of commits of size $i$ involving file $f$ in target release; \\ $D_f$ = developers of all commits involving $f$ in target release} \\
\hline

\textbf{Distinct Developer Activity Vector} ($\mathbf{V}^{(ddev)}_f$) &
Distribution of the number of distinct developers who modified file $f$ during the project's lifetime across commits of different sizes up to the target release. &
\parbox[c]{2.9cm}{\centering 
\(\mathbf{V}^{(\text{ddev})}_f[i] = \dfrac{|D^*_{f,i}|}{|D^*_f|}\)} &
\parbox[t]{6cm}{$D^*_{f,i}$ = developers of commits of size $i$ involving file $f$ during the project's lifetime up to the target release; \\ $D^*_f$ = developers of all commits involving $f$ in the project up to the target release.} \\

\hline

\textbf{Lines Added Vector} ($\mathbf{V}^{(add)}_f$) &
Distribution of lines added to $f$ across commit sizes. &
\parbox[c]{2.9cm}{\centering 
\(\mathbf{V}^{(\text{add})}_f[i] = \dfrac{|LA_{f,i}|}{|LA_f|}\)} &
\parbox[t]{6cm}{$\mathrm{LA}_{f,i}$ = lines added to $f$ in commits of size $i$;\\$\mathrm{LA}_f$ = lines added to file $f$ across all commits involving it } \\

\hline

\textbf{Lines Deleted Vector} ($\mathbf{V}^{(del)}_f$) &
Distribution of lines deleted from $f$ across commit sizes. &
\parbox[c]{2.9cm}{\centering 
\(\mathbf{V}^{(\text{del})}_f[i] = \dfrac{|LD_{f,i}|}{|LD_f|}\)} &
\parbox[t]{6cm}{$\mathrm{LD}_{f,i}$ = lines deleted to $f$ in commits of size $i$;\\$\mathrm{LD}_f$ = lines deleted to file $f$ across all commits involving it} \\

\hline

\textbf{Owner Contribution Vector} ($\mathbf{V}^{(own)}_f$) &
Owner’s contribution distribution across commit sizes. &
\parbox[c]{2.9cm}{\centering 
\(\mathbf{V}^{(\text{own})}_f[i] = \dfrac{|L^{(o)}_{f,i}|}{|L^{(o)}_f|}\)} &
\parbox[t]{6cm}{$\mathrm{L}_{f,i}^{(\mathrm{o})}$ = Lines changed by the owner in file $f$ through commits of size $i$ in the target release;
\\$\mathrm{L}_{f}^{(\mathrm{o})}$ = lines changed by the owner in all commits involving $f$ in the target release.} \\
\hline
\textbf{Owner Experience Vector} ($\mathbf{V}^{(oexp)}_f$) &
Distribution of owner's experience in the project across commit sizes . &
\parbox[c]{2.9cm}{\centering 
\(\mathbf{V}^{(\text{oexp})}_f[i] = \dfrac{|L^{(o*)}_{f,i}|}{|L^{(o*)}_f|}\)} &
\parbox[t]{6cm}{$\mathrm{L}_{f,i}^{(\mathrm{o*})}$ = lines changed by the owner of $f$ in commits of size $i$ in project lifetime till  target release;
\\$\mathrm{L}_{f}^{(\mathrm{o*})}$ = lines changed by the owner in all commits in project lifetime till target release} \\

\hline

\textbf{Developers' Experience Vector} ($\mathbf{V}^{(exp)}_f$) &
Represents the average experience of authors contributing to $f$ across the  commit sizes. &
\parbox[c]{2.9cm}{\centering 
\(\mathbf{V}^{(\text{exp})}_f[i] = \textit{G}_{\text{Mean}}\!\left\{ E_D \right\}\)} &
\parbox[t]{6cm}{
$E_D= \text{set of }  \dfrac{e(d_{j,i})}{e(d_j)}$ where,
$e(d_{j,i})$ = experience of developer $j$ who changed file $f$ through commits of size $i$;\\
$e(d_j)$ = experience of the developer $j$ who changed file $f$.
} \\

\hline

\textbf{Minor Contributors Vector} ($\mathbf{V}^{(minor)}_f$) &
Distribution of minor developers across the commit sizes . &
\parbox[c]{2.9cm}{\centering 
\(\mathbf{V}^{(\text{minor})}_f[i] = \dfrac{|M_{f,i}|}{|M_f|}\)} &
\parbox[t]{6cm}{
$M_{f,i}$ = minor developers considering commits of size $i$ involving $f$;\\$M_f$ = minor developers considering all commits involving $f$.} \\

\hline

\textbf{Size-aware change Entropy Vector (SCTR)} ($\mathbf{V}^{(sctr)}_f$) &
commit size aware Change Entropy Vector $f$  &
\parbox[c]{2.9cm}{\centering 
\(\mathbf{V}^{(\text{sctr})}_f[i] = p_i(f) . H_i\)} &
\parbox[t]{6cm}{$p_i(f)$ is the proportion of commits of size $i$ involving $f$;
\\$H_i$ is the entropy of considering commit of size $i$ across all files} \\

\hline

\textbf{Neighborhood Commit count Vector (NCOMM)} ($\mathbf{V}^{(ncomm)}_f$) &
Aggregates neighbors’ commit count vector information. &
\parbox[c]{2.9cm}{\centering $\dfrac{\sum_{n} w_{n}\mathbf{V}^{(comm)}_n}{\sum_{n} w_{n}\|\mathbf{V}^{(comm)}_n\|_1}$} &
\parbox[t]{6cm}{
$n \in \text{neighbors of } f$ in the co-change graph;\\$w_{n}$ = edge weight between $f$ and $n$ in co-change graph.
} \\

\hline

\textbf{Neighborhood Developer Activity Vector (NADEV)} ($\mathbf{V}^{(nadev)}_f$) &
Aggregates neighbors’ Developer Activity Vector information. &
\parbox[c]{2.9cm}{\centering $\dfrac{\sum_{n} w_{n}\mathbf{V}^{(adev)}_n}{\sum_{n} w_{n}\|\mathbf{V}^{(adev)}_n\|_1}$} &
\parbox[t]{6cm}{
$n \in \text{neighbors of } f$ in the co-change graph;\\$w_{n}$ = edge weight between $f$ and $n$ in co-change graph.
} \\
\hline

\textbf{Neighborhood Distinct Developer Vector (NDDEV)} ($\mathbf{V}^{(nddev)}_f$) &
Like NADEV, but using distinct author vectors of neighbors. &
\parbox[c]{2.9cm}{\centering $\dfrac{\sum_{n} w_{n}\mathbf{V}^{(ddev)}_n}{\sum_{n} w_{n}\|\mathbf{V}^{(ddev)}_n\|_1}$} &
\parbox[t]{6cm}{
$n \in \text{neighbors of } f$ in the co-change graph;\\$w_{n}$ = edge weight between $f$ and $n$ in co-change graph.
} \\

\hline

\textbf{Neighborhood Entropy Vector (NSCTR)} ($\mathbf{V}^{(nsctr)}_f$) &
Aggregating Neighbors change-entropy vector information. &
\parbox[c]{2.9cm}{\centering $\dfrac{\sum_{n} w_{n}\mathbf{V}^{(sctr)}_n}{\sum_{n} w_{n}\|\mathbf{V}^{(sctr)}_n\|_1}$} &
\parbox[t]{6cm}{
$n \in \text{neighbors of } f$ in the co-change graph;\\$w_{n}$ = edge weight between $f$ and $n$ in co-change graph.
} \\

\hline
\end{tabular}
\end{table*}
RQ1 investigates whether size-aware vector process metrics provide improvements in defect classification over their scalar counterparts and whether they complement popular product metrics more effectively. To address this, we first extracted the commits corresponding to each release of every project in our dataset. The projects were cloned from their respective GitHub repositories, and commits were retrieved using tag information and the \texttt{git rev-list} command, following the procedure adopted in previous studies \cite{wen2018well,majumder2022revisiting}.
\par Recent studies have cautioned that range-based strategies are more accurate than time-based ones for assigning commits to releases \cite{do2021assessing,pinto2023assignment}. Hence, we employed a range-based approach to minimize errors in commit-to-release mapping. After obtaining the release-wise commit data, we computed all conventional process metrics as defined by Rahman and Devanbu \cite{rahman2013and} for the source code files of all releases and for all projects included in our dataset. This set of 14 widely used process metrics, as defined by Rahman and Devanbu \cite{rahman2013and}, has been adopted in replication and benchmarking studies on defect prediction \cite{majumder2022revisiting} and is commonly used for evaluating process-based defect predictors.
Following recent methodological recommendations \cite{esposito2023uncovering,liu2024unveiling}, we excluded untouched modules—files that were never modified during the studied releases—as their inclusion can bias the results. For such modules, product metrics remain static while process metrics lack meaningful values. Consequently, our dataset includes only those source code files that were modified at least once during the analyzed releases.
\par
For each source code file modified in release $R_j$, we obtain its product metric values and post-release defect label from the dataset corpus of Yatish et~al.\ \ In contrast, the conventional process metrics defined by Rahman and Devanbu \cite{rahman2013and}, along with their size-aware vector counterparts, are computed by us. Because commits that modify more than 100 files are excluded, each process-metric vector has 100 components, where the $i^{\text{th}}$ component aggregates the contribution from commits of size $i$. As an example, for the scalar \textit{Commit Count} metric, we construct a \textit{Commit Count Vector} for each file $f$, where the $i^{\text{th}}$ component represents the proportion of commits that modified $f$ and had a size of $i$. For brevity, we do not enumerate the product metrics or the scalar process metrics here; the scalar process metrics are detailed in Rahman and Devanbu \cite{rahman2013and}, and the product metrics used in this study are described by Yatish et~al.\ \cite{yatish2019mining}. Table \ref{tab:vectorprocess4} illustrates how each vector process metric is derived from its corresponding scalar counterpart. Since each of the 14 scalar process metrics is expanded into a 100-dimensional vector form, this yields a total of 1,400 vector process metrics. In summary, our feature set comprises 54 product metrics, 14 scalar process metrics, and 1,400 vector process metrics used for classification. 
\subsubsection{Evaluation Protocol}
For each project release, we collect all files changed in that release and obtain their product metrics, scalar process metrics, and size-aware vector process metrics as described above. We also extract each file’s post-release defect label. However, not all metrics were used to train the classifiers. Before model training, we performed feature selection using the HSIC Lasso method. The HSIC Lasso requires specifying the number of top features ($k$) to be selected. We experimented with different values of $k$ and observed that the best performance was obtained when $k=40$. Consequently, we selected the top 40 features from each feature set combination for all subsequent experiments.
\par We evaluate performance using bootstrap validation with 100 out-of-sample resamples, per project and per classifier. Across 9 projects and 5 classifiers, this yields 45 project--classifier configurations. This is in accordance with past studies \cite{yatish2019mining}. For each configuration, we train and test two feature sets: (i) Product + Scalar Process (PR+SP) and (ii) Product + Vector Process (PR+VP). If PR+VP outperforms PR+SP, we infer that vector process metrics add value for defect classification.
\par To assess performance comprehensively, we report F1, AUROC, MCC, Brier score, and area under the precision--recall curve (AUPRC). We do not report accuracy, since defect datasets are typically imbalanced and accuracy is not appropriate in such settings \cite{song2018comprehensive,moussa2022use}. F1 balances precision and recall for the minority (buggy) class; MCC summarizes all four confusion matrix terms and is robust under imbalance \cite{yao2020assessing}; AUROC measures ranking ability across thresholds; AUPRC emphasizes performance on the positive class; and the Brier score evaluates the calibration of predicted probabilities. These measures are widely used in defect prediction studies \cite{gong2021revisiting,yatish2019mining}.
\subsection{Experimental Setup to Answer RQ2}
\subsubsection{Details of the Metrics}
Along with the product metrics, scalar process metrics, and vector process metrics discussed in Section~\ref{RQ2_setup}, a fourth category of metrics used in this study is derived from the \textit{co-change graph}. Traditionally, a simple co-change graph connects two source code files if they are modified together in the same commit. However, such pairwise graphs fail to encode commit-size information—i.e., whether the files co-occurred in small or large commits.

To overcome this limitation, we construct a \textit{hyper co-change graph} from the commit history, where nodes represent source code entities and each hyperedge links all files modified together in a commit. Unlike pairwise networks, where edges connect only two nodes, hypergraphs allow hyperedges connecting multiple nodes simultaneously, thereby preserving the group-wise semantics of co-changes.

\begin{figure}[htbp]
    \centering
    \begin{adjustbox}{max width=\columnwidth}
        \begin{subfigure}[b]{0.24\textwidth}
            \centering
            \includegraphics[width=\linewidth]{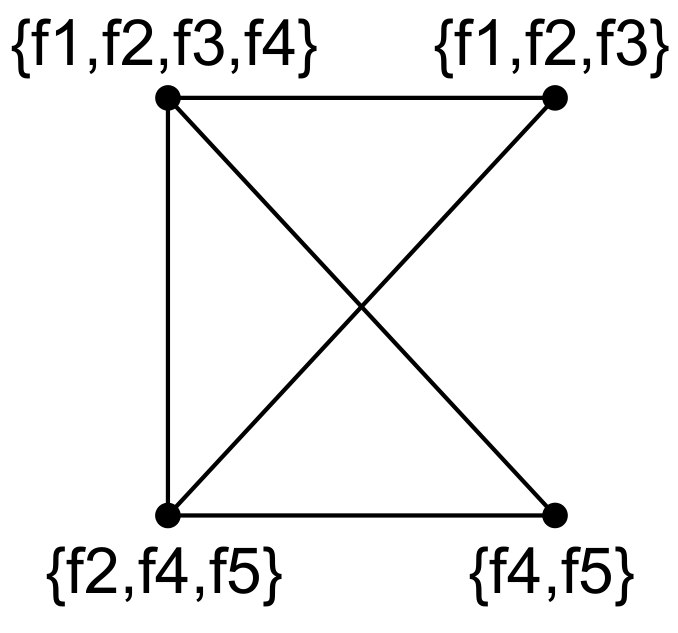}
            \caption{Line graph}
            \label{fig:line_graph}
        \end{subfigure}\hfill
        \begin{subfigure}[b]{0.24\textwidth}
            \centering
            \includegraphics[width=\linewidth]{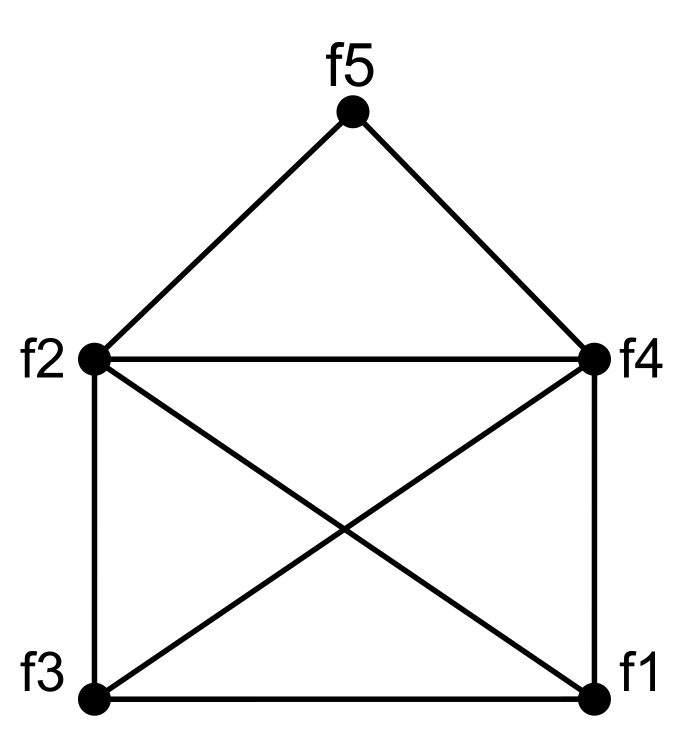}
            \caption{Pairwise co-change graph}
            \label{fig:normal_graph}
        \end{subfigure}\hfill
        \begin{subfigure}[b]{0.24\textwidth}
            \centering
            \includegraphics[width=\linewidth]{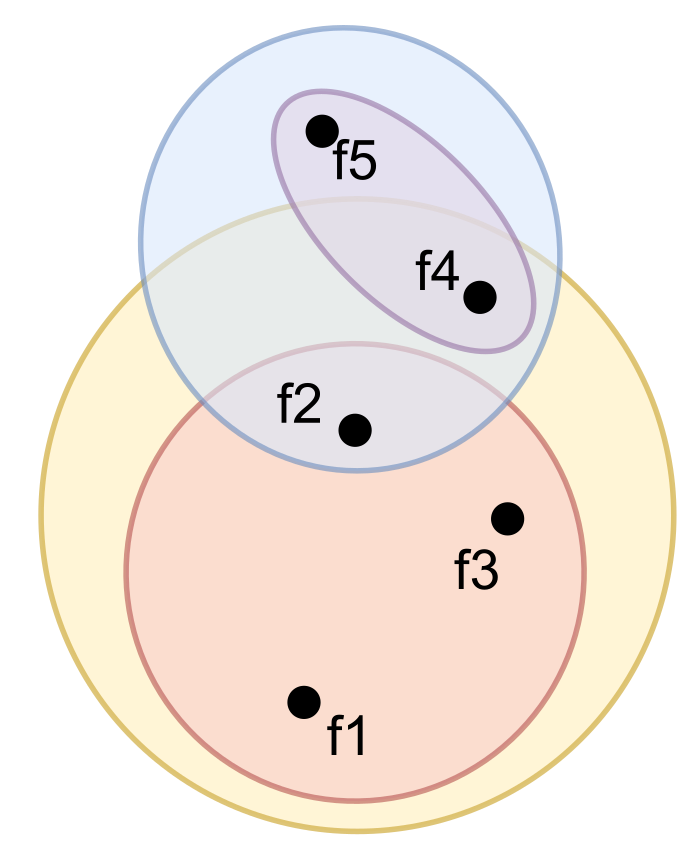}
            \caption{Hypergraph (Set 1)}
            \label{fig:hypergraph_1}
        \end{subfigure}\hfill
        \begin{subfigure}[b]{0.24\textwidth}
            \centering
            \includegraphics[width=\linewidth]{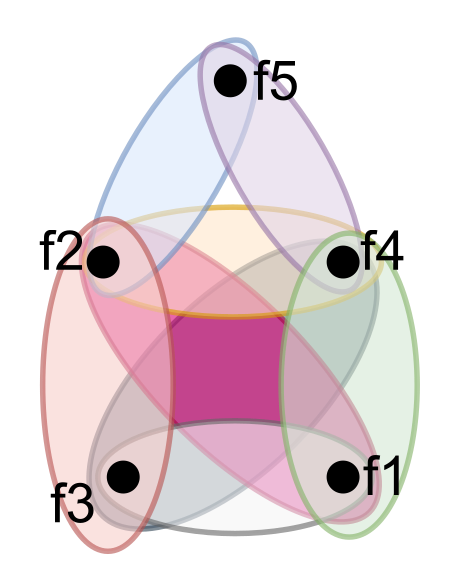}
            \caption{Hypergraph (Set 2)}
            \label{fig:hypergraph_2}
        \end{subfigure}
    \end{adjustbox}
    \caption{Hyper co-change vs. pairwise co-change networks in capturing commit size. Pairwise graphs collapse group edits into binary links, obscuring commit size, while hypergraphs preserve the full change set via hyperedges.}
    \label{fig:hyper_cochange_graph}
\end{figure}

Table~\ref{tab:sample_change_sets} presents a sample change history, showing the files and the commits \textit{(C1,..,C8)} in which they were modified. In a simple pairwise co-change graph, two file nodes are connected if they were changed in the same commit, whereas in a hyper co-change graph, all files changed together in a commit form a single hyperedge.
As illustrated in Figure~\ref{fig:hyper_cochange_graph} and Table~\ref{tab:sample_change_sets}, two change sets of markedly different nature can produce identical pairwise co-change graphs (Figure~\ref{fig:normal_graph}) but distinct hypergraphs (Figures~\ref{fig:hypergraph_1} and~\ref{fig:hypergraph_2}) that retain commit-size information. Consequently, centrality measures on pairwise co-change graphs cannot distinguish a file’s importance arising from participation in large versus small commits—an important distinction naturally captured by the hypergraph formulation.

\begin{table}[htbp]
\centering
\small
\resizebox{0.48\textwidth}{!}{%
\begin{tabular}{|l|l|l|l|l|l|}
\hline
\textbf{Files} & \textbf{F1} & \textbf{F2} & \textbf{F3} & \textbf{F4} & \textbf{F5} \\
\hline
\textbf{Change Set 1} & \textit{C1, C4} & \textit{C1, C2, C4} & \textit{C1, C4} & \textit{C1, C2, C3} & \textit{C2, C3} \\
\textbf{Change Set 2} & \textit{C1, C5, C6} & \textit{C1, C7, C8} & \textit{C3, C4} & \textit{C2, C3, C5, C7} & \textit{C2, C4, C6, C8} \\
\hline
\end{tabular}%
}
\caption{Example change sets used for the network comparison in Figure~\ref{fig:hyper_cochange_graph}.}
\label{tab:sample_change_sets}
\end{table}

For node importance measures in hypergraph we adopt the approach proposed by Kovalenko et al.~\cite{kovalenko2022vector}, which introduces \emph{vector centralities} for hypergraphs, generalizing classical centrality measures to higher-order interactions.

Vector centrality extends standard SNA metrics—such as degree, betweenness, closeness, and eigenvector centrality—to hypergraphs, where interactions can involve more than two nodes simultaneously. Let $G = (N, \mathcal{M})$ be an undirected hypergraph, where $N$ is the set of nodes and $\mathcal{M} \subseteq 2^N$ is the set of hyperedges (non-empty subsets of $N$). Let $H = \max_{e \in \mathcal{M}} |e|$ be the maximum cardinality (or order) of any hyperedge.

To compute vector centralities, we construct the \emph{line graph} (also called the \emph{hyperedge adjacency graph}) of the hypergraph $G$, denoted \( I(G) \). In this graph, each node represents a hyperedge \( e \in \mathcal{M} \), and an undirected edge is added between two hyperedges \( e, e' \in \mathcal{M} \) if they share at least one common node (i.e., \( e \cap e' \neq \emptyset \)). This line graph encodes the pairwise overlap structure among hyperedges, enabling the application of classical centrality measures to hyperedges themselves. In our example, for Change Set 1 shown in Table~\ref{tab:sample_change_sets} and the corresponding hypergraph depicted in Figure~\ref{fig:hypergraph_1}, the associated line graph is shown in Figure~\ref{fig:line_graph}.

Let \( x(e) \) denote the centrality score of a hyperedge \( e \in \mathcal{M} \) as computed on \( I(G) \) using a classical centrality measure (e.g., eigenvector, closeness, betweenness, or degree centrality). The $l$-th component of the vector centrality for node $j \in N$ is defined as:

\begin{equation}
x_{j,l} = \frac{1}{l} \sum_{\substack{e \in \mathcal{M} \\ |e| = l \\ j \in e}} x(e),
\end{equation}

for \( l = 2, 3, \ldots, H \). This expression quantifies the importance of a node within the hypergraph by considering its participation in hyperedges of size \( l \). Specifically, it captures how central a node is when interacting with other nodes through group associations of order \( l \). In the context of vector centrality, this component-level view allows us to disentangle a node's role across different interaction orders, rather than collapsing all contributions into a single scalar score. By normalizing the contribution of each hyperedge by its size, the measure avoids overemphasizing large hyperedges and provides a more balanced perspective on the node's involvement in group-level interactions of a specific order. It has been shown that $\sum_{j \in N} \| \vec{x}_j \|_1=1$ making it a properly normalized measure.
\par It should be noted that we compute vector centralities for each node corresponding to the global centrality measures commonly used in pairwise networks—namely, degree, betweenness, closeness, and eigenvector centralities. Since we filter out commits that modify more than 100 files, the maximum number of components in the vector centrality of a node for a given centrality type (e.g., betweenness) is 100. Considering all four centrality measures, the hypergraph-based vector centralities yield a total of 400 features for each node (source code file).
\subsubsection{Evaluation Protocol}
To answer RQ2, we conduct experiments across all project–classifier pairs using the same machine learning pipeline employed for RQ1. The objective here is to examine whether incorporating vector centrality further enhances defect prediction performance. For this purpose, we evaluate classifiers using the metric set \textit{Product + Vector Process Metrics + Vector Centrality Scores} (PR+VP+VC), applying the same evaluation metrics as in RQ1. 
\par Our primary focus is to determine whether the performance gain achieved by PR+VP+VC over PR+SP (scalar process metrics) exceeds the improvement observed when using PR+VP over PR+SP. To assess this, we first compute how often (out of 45 project–classifier combinations) each metric set—PR+SP, PR+VP, and PR+VP+VC—achieves the best performance for each evaluation measure (e.g., AUROC, MCC). However, simple frequency counts do not establish statistical confidence. Therefore, following the past studies\cite{gong2021revisiting}, we apply the Friedman test  to determine whether the performance differences among the metric sets are statistically significant, followed by the post-hoc Nemenyi test to identify pairwise rankings. This analysis is conducted using the dataset-classifier pair results for all the metrics considered in this study. Performing statistical validation across dataset-classifier pairs helps eliminate biases in the conclusions drawn from machine learning experiments, as also emphasized in recent studies within the machine learning domain \cite{tomani2023beyond,singhalfoundation}.
\par In practice, the ability of defect classifiers to prioritize the most defect-prone files is often more relevant. Hence, we also evaluate the top-$k$ ranking performance by ordering source code files according to their predicted probability of being defective and analyzing the resulting Precision@k and Recall@k curves.
\section{Results and discussion}
\subsection{Answer to RQ1}
\begin{figure*}[htbp]
    \centering
    \begin{adjustbox}{max width=\linewidth}
        \includegraphics{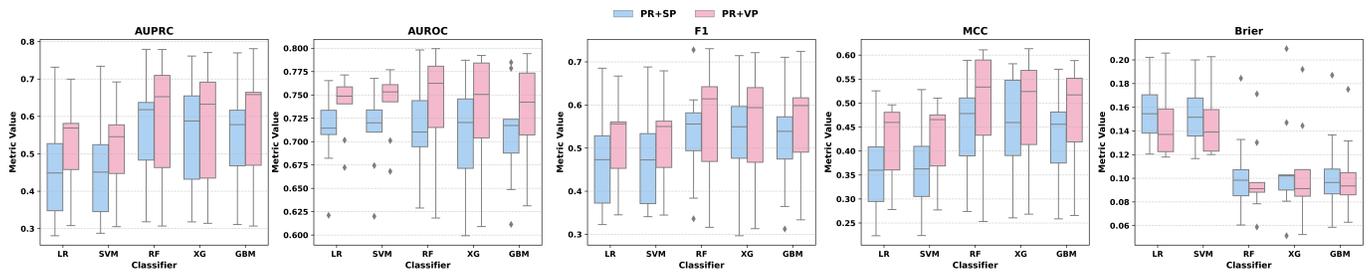}
    \end{adjustbox}
    \caption{Absolute performance of PR+SP and PR+VP feature combinations across multiple metrics}
    \label{fig:box_PT_PV}
\end{figure*}
To answer RQ1, we trained our classifiers using two feature sets: Product + Scalar Process (PR+SP) and Product + Vector Process (PR+VP) metrics. The experiments were conducted across 45 dataset--classifier pairs (9 projects $\times$ 5 classifiers), and the aggregated results across projects are shown as box plots in Fig.~\ref{fig:box_PT_PV}. The figure shows that PR+VP consistently outperforms PR+SP across all classifiers and evaluation metrics. The distributions of PR+VP scores are shifted in the favorable direction, exhibiting higher medians for AUROC, AUPRC, F1, and MCC, and lower medians for Brier scores, indicating systematic performance gains rather than improvements restricted to a few outliers. These results confirm that incorporating commit-size semantics into process metrics enhances prediction accuracy, ranking ability, and calibration of defect prediction models.
\par
The median AUPRC improves by 5.6\%--26.7\% (maximum with Logistic Regression), suggesting a higher proportion of true defects retrieved among top-ranked files under class imbalance. AUROC increases by 3.5\%--7.3\% (maximum with Random Forest), reflecting improved discrimination between defective and clean files across thresholds. F1 rises by 8.0\%--17.6\% (maximum with Logistic Regression), demonstrating a better precision--recall balance. MCC improves by 11.5\%--28.3\% (maximum with SVM), indicating stronger agreement between predictions and actual outcomes. The Brier score decreases by 3.0\%--11.3\% (maximum with Logistic Regression), confirming better probability calibration and reliability for risk-based decisions.
\par
Although no single classifier dominates across all metrics, the consistent distributional shifts and significant median gains clearly demonstrate that integrating commit-size semantics through vector process metrics (PR+VP) produces more discriminative, balanced, and well-calibrated defect prediction models than scalar process metrics (PR+SP). In a nutshell, the answer to RQ1 is as follows: \\
\begin{center}
\fbox{\begin{minipage}{\dimexpr\columnwidth-2\fboxsep-2\fboxrule}
\textbf{Answer to RQ1:} \textit{Incorporating commit-size semantics through vector process metrics (PR+VP) significantly enhances the effectiveness of product metrics compared to scalar process metrics (PR+SP). Median improvements range between 3.5\%--7.3\% for AUROC, 5.6\%--26.7\% for AUPRC, 8.0\%--17.6\% for F1, and 11.5\%--28.3\% for MCC, while Brier scores decrease by 3.0\%--11.3\%. These consistent gains confirm that commit-size–aware process representations yield more discriminative, better-calibrated, and balanced defect prediction models.}
\end{minipage}}
\end{center}
\subsection{Answer to RQ2}
\begin{figure*}[htbp]
    \centering
    \begin{adjustbox}{max width=\linewidth}
        \includegraphics{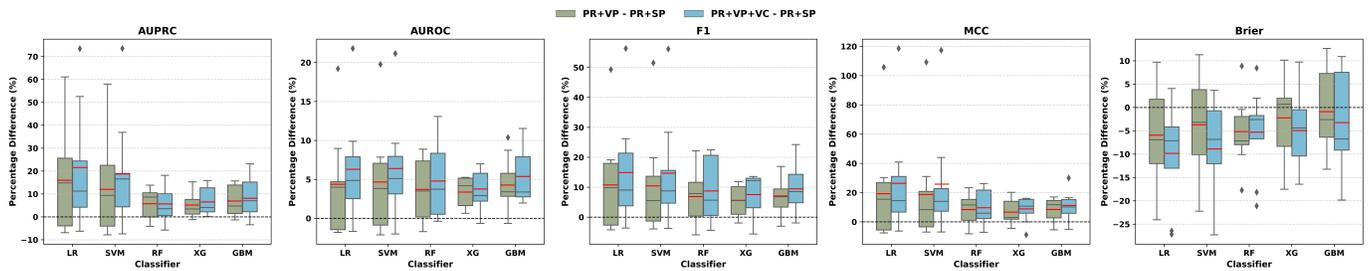}
    \end{adjustbox}
    \caption{Comparison of percentage improvements across performance metrics from PR+SP to PR+VP, and from PR+SP to PR+VP+VC}
    \label{fig:box_difference}
\end{figure*}

To answer RQ2, we trained classifiers using the combined feature set PR+VP+VC, which integrates product metrics, vector process metrics, and vector centrality measures. To assess whether incorporating vector centrality features yields additional benefits beyond PR+VP, we computed the evaluation metrics (e.g., AUROC) for each dataset--classifier pair and analyzed the percentage improvements relative to the baseline PR+SP configuration. Specifically, we compared the improvements achieved by PR+VP over PR+SP and those achieved by PR+VP+VC over PR+SP.  
\par  
The results are presented in Figure~\ref{fig:box_difference}. Each box plot depicts the percentage improvement over the PR+scalar process baseline (PR+SP) across projects for each classifier, comparing two configurations: PR+VP and PR+VP+VC. The gray line within each box represents the median (Q2), while the red line indicates the mean. For metrics such as AUROC, F1, AUPRC and MCC---where higher values imply better performance---the median and mean percentage gains of  PR+VP+VC  consistently exceed those of  PR+VP. In several classifiers, the entire interquartile range (IQR) of the  PR+VP+VC  boxes shifts upward, indicating both higher and more stable performance improvements across projects. This clearly demonstrates that incorporating vector centrality features further enhances model performance on both rank- and threshold-based measures.  
\par  
For the Brier score, where lower values indicate better calibration, the boxes are centered around negative percentage differences relative to  PR+SP. Here,  PR+VP+VC  exhibits more negative median and mean values than  PR+VP, reflecting a greater reduction in Brier score and consequently better model calibration.
\par 
As evident from Figures~\ref{fig:box_PT_PV} and~\ref{fig:box_difference}, the PR+VP+VC configuration consistently outperforms both PR+VP and PR+SP, confirming the overall utility of size-aware process metrics. Since the numerical differences among these feature sets appear moderate, it is essential to assess whether these improvements are statistically significant. To this end, we applied the Friedman test across all dataset–classifier pairs for the three feature sets (PR+SP, PR+VP, and PR+VP+VC). The results yielded a p-value $<0.05$, indicating significant differences among their performance scores.
\begin{figure}[htbp]
    \centering
    \begin{adjustbox}{max width=\columnwidth} 
        \begin{minipage}{\columnwidth}
            \centering
            \begin{subfigure}[b]{\columnwidth}
                \centering
                \includegraphics[width=\linewidth]{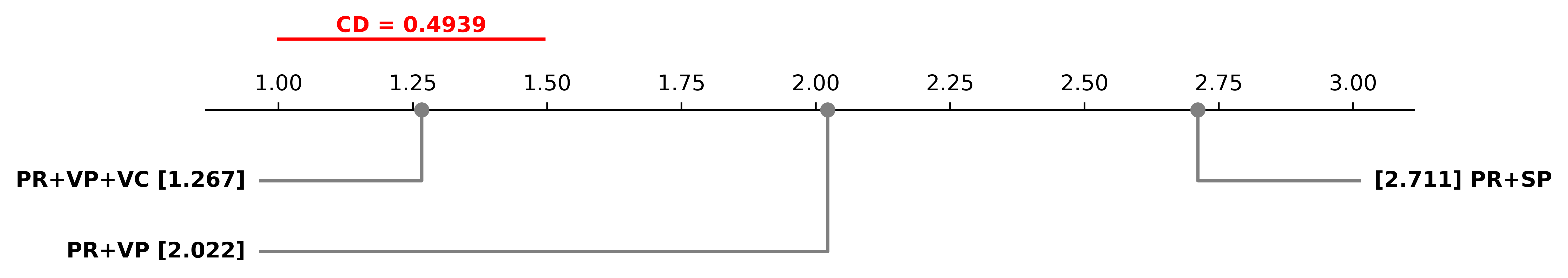}
                \caption{AUROC CD plot}
                \label{fig:auroc_cd}
            \end{subfigure}
            
            \begin{subfigure}[b]{\columnwidth}
                \centering
                \includegraphics[width=\linewidth]{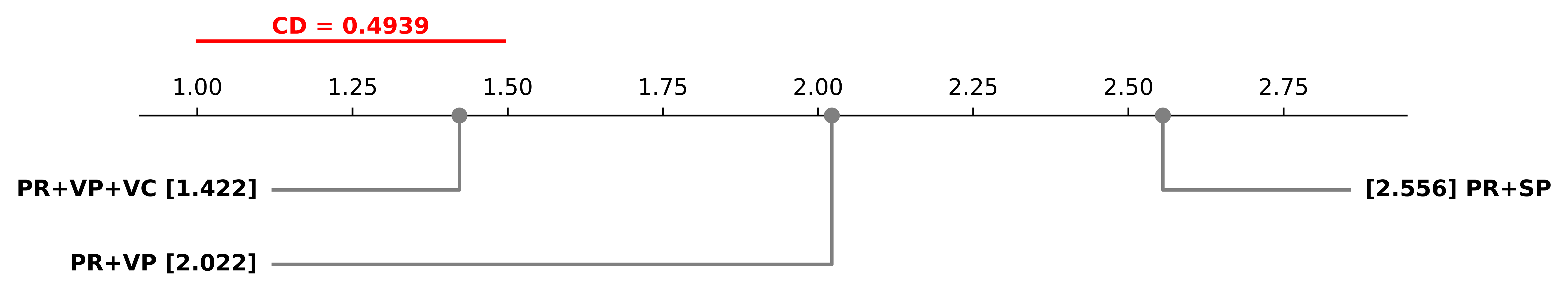}
                \caption{MCC CD plot}
                \label{fig:mcc_cd}
            \end{subfigure}
            
            \begin{subfigure}[b]{\columnwidth}
                \centering
                \includegraphics[width=\linewidth]{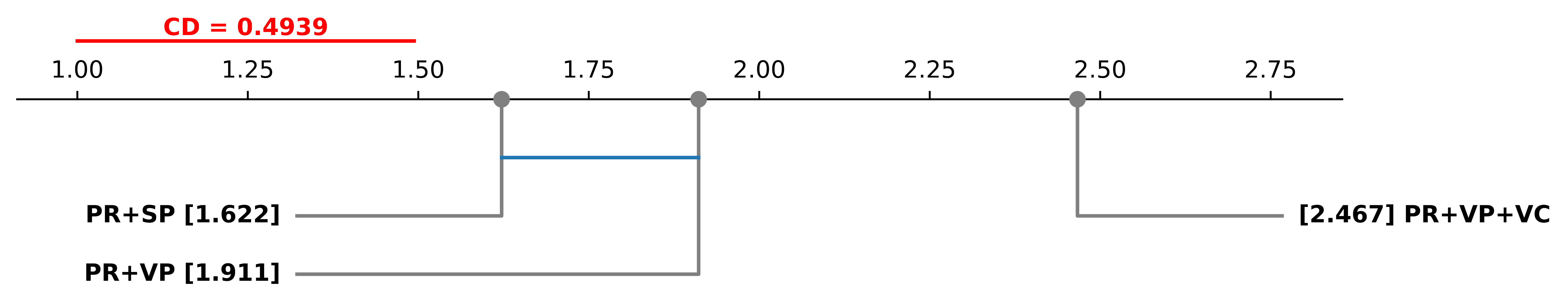}
                \caption{Brier CD plot}
                \label{fig:brier_cd}
            \end{subfigure}
            
            \begin{subfigure}[b]{\columnwidth}
                \centering
                \includegraphics[width=\linewidth]{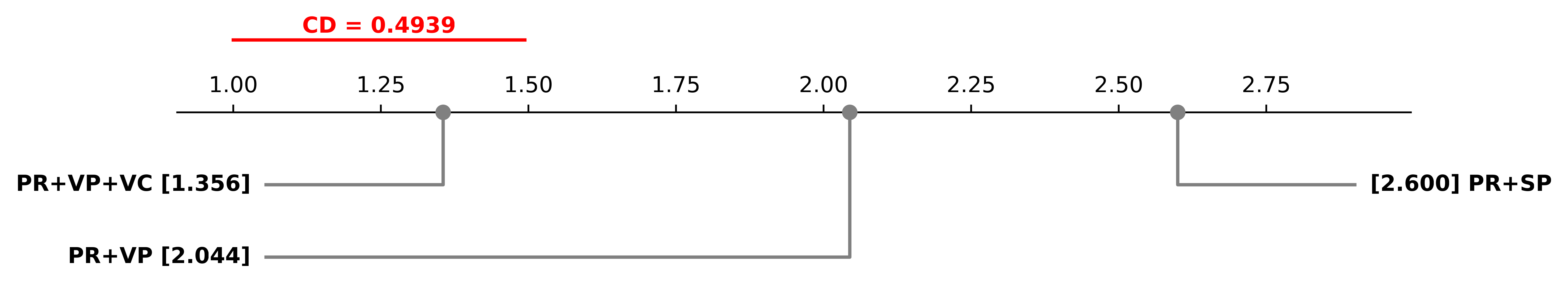}
                \caption{F1 CD plot}
                \label{fig:f1_cd}
            \end{subfigure}
            
            \begin{subfigure}[b]{\columnwidth}
                \centering
                \includegraphics[width=\linewidth]{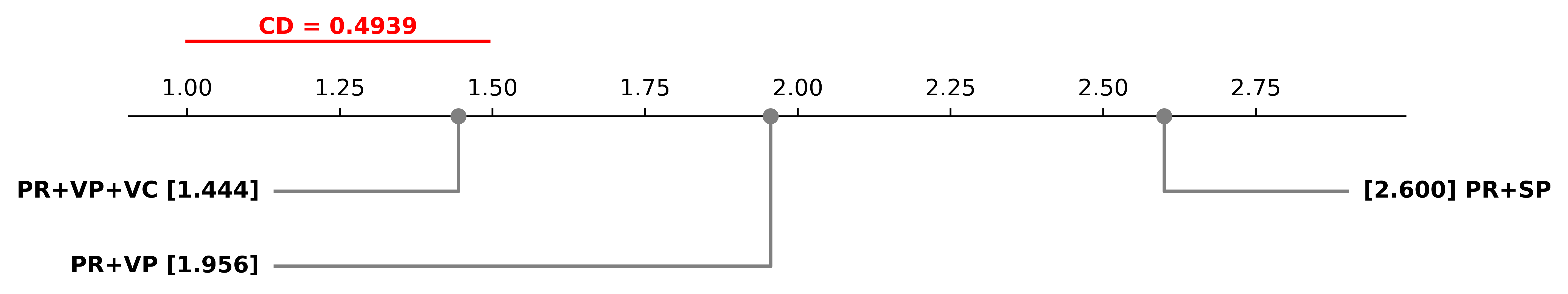}
                \caption{AUPRC CD plot}
                \label{fig:auc_cd}
            \end{subfigure}
        \end{minipage}
    \end{adjustbox}
    \caption{Critical Difference plots showing the results of the post-hoc Nemenyi test}
    \label{fig:five_vertical}
\end{figure}
\par 
To further identify which feature set performs best, we conducted a post-hoc Nemenyi test. The results, presented in Figure~\ref{fig:five_vertical} reveal that except for the Brier score, the improvements of PR+VP over PR+SP are statistically significant across all other four metrics. Moreover, the mean rank differences between PR+VP+VC and both PR+VP and PR+SP exceed the critical distance in the CD plots, confirming that the inclusion of vector centrality features provides statistically significant gains across all five evaluation metrics. Overall, the Friedman–Nemenyi tests demonstrate that PR+VP+VC achieves the highest ranks for all classification performance measures, outperforming both baselines. 
\par
\begin{figure}[htbp]
    \centering
    \begin{adjustbox}{max width=\linewidth}
    \includegraphics{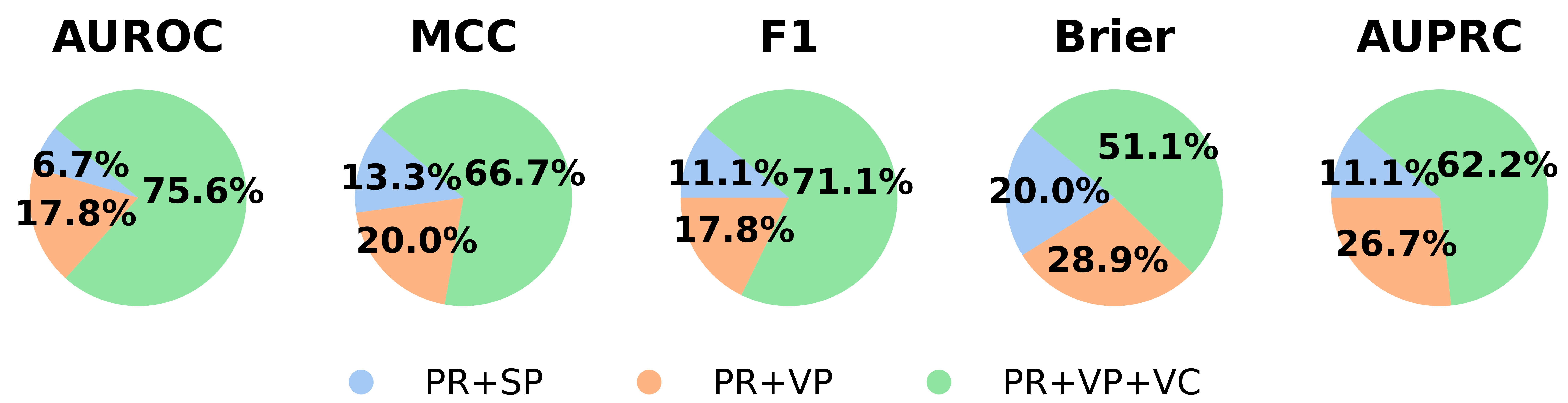}
    \end{adjustbox}
    \caption{Pie chart showing the percentage of times each feature combination yields the best performance}
    \label{fig:piechart}
\end{figure}

 
The pie chart in Figure~\ref{fig:piechart} summarizes the frequency with which each feature set achieved the best performance across 45 dataset–classifier combinations. PR+VP+VC achieved the highest proportion of best performances—75.6\% in AUROC, 66.7\% in MCC, 71.1\% in F1, 62.2\% in AUPRC, and lowest Brier scores in 51.1\% of cases. The clear and significant  wider gap between PR+VP+VC and PR+SP in the CD plots, along with its consistently lower Brier scores, underscores the added confidence and calibration achieved by incorporating vector centrality features.
\par
\begin{figure*}[htbp]
    \centering
    \begin{adjustbox}{max width=\linewidth}
        \includegraphics{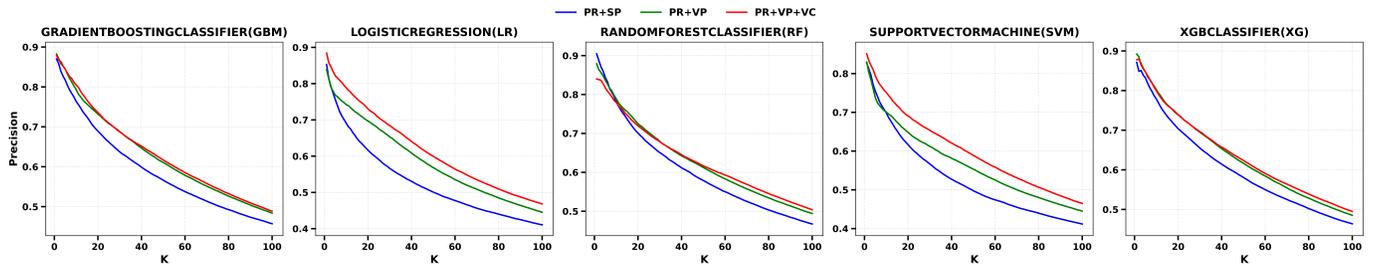}
    \end{adjustbox}
    \caption{Mean Precision at k for different feature combinations across classifiers}
    \label{fig:precision@k}
\end{figure*}
\begin{figure*}[htbp]
    \centering
    \begin{adjustbox}{max width=\linewidth}
      \includegraphics{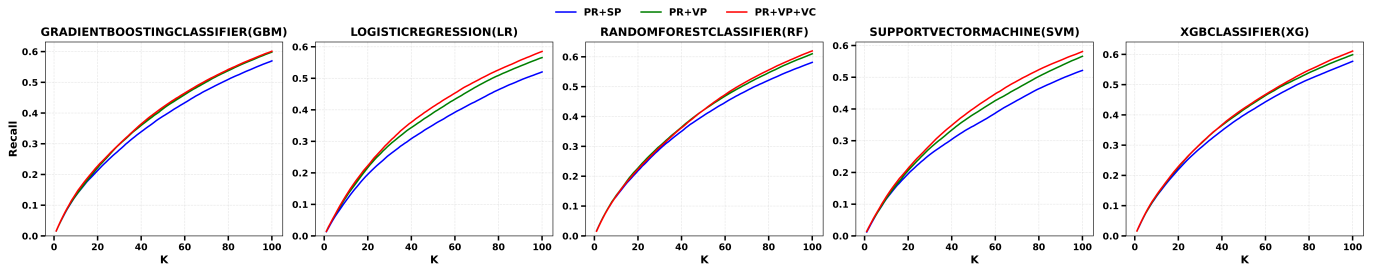}
    \end{adjustbox}
    \caption{Mean Recall at k for different feature combinations across classifiers}
    \label{fig:recall@k}
\end{figure*}

Since one of the key objectives in defect prediction is to rank source code files according to their defect proneness, we also evaluated the models from a ranking perspective. We assessed the top-$k$ performance of all three feature sets using mean \textit{precision@k} (mean across the projects) and \textit{recall@k} for $k = 1$ to $100$, as shown in Figures~\ref{fig:precision@k} and~\ref{fig:recall@k}. The results indicate that although PR+VP+VC performs best overall, its \textit{precision@k} and \textit{recall@k} values are only marginally higher than those of PR+VP. However, both size-aware configurations (PR+VP and PR+VP+VC) substantially outperform PR+SP across all $k$ values. 

\par 
In summary, the results confirm that incorporating vector centrality features over the hyper co-change graph significantly improves defect classification performance and ranking quality. Among all three feature sets—PR+SP, PR+VP, and PR+VP+VC—the PR+VP+VC configuration consistently delivers the best and most reliable outcomes.
\begin{center}
\fbox{\begin{minipage}{\dimexpr\columnwidth-2\fboxsep-2\fboxrule}
\textbf{Answer to RQ2:} \textit{Incorporating vector centrality features over the hyper co-change graph (PR+VP+VC) yields statistically significant improvements over both PR+VP and PR+SP. The Friedman test  and post-hoc Nemenyi test confirm that PR+VP+VC consistently achieves the best mean ranks across all five evaluation metrics. Across all 45 dataset–classifier pairs, PR+VP+VC achieves the highest frequency of best performance—75.6\% in AUROC, 66.7\% in MCC, 71.1\% in F1, 62.2\% in AUPRC, and lower Brier scores in 51.1\% of cases. Both PR+VP and PR+VP+VC also outperform PR+SP on ranking-based metrics (\textit{precision@k}, \textit{recall@k}) for all $k \in [1,100]$. Overall, the addition of vector centrality features enhances both discriminative and calibration capabilities of defect classifiers, establishing PR+VP+VC as the most effective and statistically superior feature set.}
\end{minipage}}
\end{center}

\section{Threats to Validity}
There are several potential threats to the validity of this study. 
First, although the experiments were conducted on a widely used dataset in software engineering research, it includes only Java projects, which may limit the generalizability of the results. 
Second, we relied on a single dataset to evaluate our approach; experiments on additional and more diverse datasets would further strengthen the generalizability of the findings. 
Finally, the combined feature set comprising product (PR), vector process (VP), and vector centrality (VC) metrics results in a high-dimensional space of approximately 1,800 features. 
Although we employ a scalable feature selection technique (HSIC Lasso) and train classifiers on the top 40 selected features, the overall model remains computationally expensive. 
\section{Conclusion}
This study proposed two novel approaches to enhance file-level software defect prediction by integrating commit-size semantics and higher-order co-change relationships into process metrics and network-based representations. First, we redefined conventional scalar process metrics into \textit{commit-size–aware vector process metrics} that capture how a file’s change behavior is distributed across commits of different magnitudes. Extensive experiments on nine open-source projects demonstrated that these vectorized metrics, when combined with product metrics, substantially improve prediction accuracy, calibration, and ranking performance over traditional scalar process metrics. 
Second, we modeled co-change relationships among files as a \textit{hyper co-change network} and employed \textit{vector centrality} measures to quantify file importance across commits of varying sizes. Integrating these hypergraph-based centralities with product and vector process metrics yielded statistically significant performance gains across multiple classifiers and evaluation metrics, as confirmed by Friedman and Nemenyi tests. 
Overall, the findings establish that commit-size semantics—captured through vector process metrics and hypergraph vector centralities—provide a richer and more discriminative characterization of software change behavior. These insights open avenues for further research on size-aware and higher-order representations in predictive software analytics.

\section*{Data Availability}
To promote transparency and reproducibility, we provide a replication package\footnote{https://github.com/unpaidresearcher/SANER2026} containing the code, dataset, and detailed instructions to replicate the experiments.

\bibliographystyle{IEEEtran}
\bibliography{References}

@article{vsikic2021improving,
  title={Improving software defect prediction by aggregated change metrics},
  author={{\v{S}}iki{\'c}, Lucija and Afri{\'c}, Petar and Kurdija, Adrian Satja and {\v{S}}Ili{\'c}, Marin},
  journal={IEEE access},
  volume={9},
  pages={19391--19411},
  year={2021},
  publisher={IEEE}
}

@inproceedings{yan2017file,
  title={File-level defect prediction: Unsupervised vs. supervised models},
  author={Yan, Meng and Fang, Yicheng and Lo, David and Xia, Xin and Zhang, Xiaohong},
  booktitle={2017 ACM/IEEE International Symposium on Empirical Software Engineering and Measurement (ESEM)},
  pages={344--353},
  year={2017},
  organization={IEEE}
}

@article{song2018comprehensive,
  title={A comprehensive investigation of the role of imbalanced learning for software defect prediction},
  author={Song, Qinbao and Guo, Yuchen and Shepperd, Martin},
  journal={IEEE Transactions on Software Engineering},
  volume={45},
  number={12},
  pages={1253--1269},
  year={2018},
  publisher={IEEE}
}

@article{pinto2023assignment,
  title={On the assignment of commits to releases},
  author={Pinto, Felipe Curty do Rego and Murta, Leonardo Gresta Paulino},
  journal={Empirical Software Engineering},
  volume={28},
  number={2},
  pages={32},
  year={2023},
  publisher={Springer}
}

@inproceedings{do2021assessing,
  title={Assessing time-based and range-based strategies for commit assignment to releases},
  author={do Rego Pinto, Felipe Curty and Costa, Bruno and Murta, Leonardo},
  booktitle={2021 IEEE international conference on software analysis, evolution and reengineering (SANER)},
  pages={142--153},
  year={2021},
  organization={IEEE}
}

@inproceedings{jiarpakdee2018autospearman,
  title={AutoSpearman: Automatically Mitigating Correlated Software Metrics for Interpreting Defect Models},
  author={Jiarpakdee, Jirayus and Tantithamthavorn, Chakkrit and Treude, Christoph},
  booktitle={2018 IEEE International Conference on Software Maintenance and Evolution (ICSME)},
  pages={92--103},
  year={2018},
  organization={IEEE}
}

@article{jiarpakdee2019impact,
  title={The impact of correlated metrics on the interpretation of defect models},
  author={Jiarpakdee, Jirayus and Tantithamthavorn, Chakkrit and Hassan, Ahmed E},
  journal={IEEE Transactions on Software Engineering},
  volume={47},
  number={2},
  pages={320--331},
  year={2019},
  publisher={IEEE}
}

@inproceedings{esposito2023uncovering,
  title={Uncovering the hidden risks: The importance of predicting bugginess in untouched methods},
  author={Esposito, Matteo and Falessi, Davide},
  booktitle={2023 IEEE 23rd International Working Conference on Source Code Analysis and Manipulation (SCAM)},
  pages={277--282},
  year={2023},
  organization={IEEE}
}

@article{liu2024unveiling,
  title={Unveiling the impact of unchanged modules across versions on the evaluation of within-project defect prediction models},
  author={Liu, Xutong and Zhou, Yufei and Lu, Zeyu and Mei, Yuanqing and Yang, Yibiao and Qian, Junyan and Zhou, Yuming},
  journal={Journal of Software: Evolution and Process},
  volume={36},
  number={12},
  pages={e2715},
  year={2024},
  publisher={Wiley Online Library}
}

@article{wattanakriengkrai2020predicting,
  title={Predicting defective lines using a model-agnostic technique},
  author={Wattanakriengkrai, Supatsara and Thongtanunam, Patanamon and Tantithamthavorn, Chakkrit and Hata, Hideaki and Matsumoto, Kenichi},
  journal={IEEE Transactions on Software Engineering},
  volume={48},
  number={5},
  pages={1480--1496},
  year={2020},
  publisher={IEEE}
}

@inproceedings{lee2023empirical,
  title={An empirical comparison of model-agnostic techniques for defect prediction models},
  author={Lee, Gichan and Lee, Scott Uk-Jin},
  booktitle={2023 IEEE International Conference on Software Analysis, Evolution and Reengineering (SANER)},
  pages={179--189},
  year={2023},
  organization={IEEE}
}

@article{jiarpakdee2020empirical,
  title={An empirical study of model-agnostic techniques for defect prediction models},
  author={Jiarpakdee, Jirayus and Tantithamthavorn, Chakkrit Kla and Dam, Hoa Khanh and Grundy, John},
  journal={IEEE Transactions on Software Engineering},
  volume={48},
  number={1},
  pages={166--185},
  year={2020},
  publisher={IEEE}
}

@inproceedings{li2022robust,
  title={Robust learning of deep predictive models from noisy and imbalanced software engineering datasets},
  author={Li, Zhong and Pan, Minxue and Pei, Yu and Zhang, Tian and Wang, Linzhang and Li, Xuandong},
  booktitle={Proceedings of the 37th IEEE/ACM International Conference on Automated Software Engineering},
  pages={1--13},
  year={2022}
}

@inproceedings{moussa2022use,
  title={On the use of evaluation measures for defect prediction studies},
  author={Moussa, Rebecca and Sarro, Federica},
  booktitle={Proceedings of the 31st ACM SIGSOFT International Symposium on Software Testing and Analysis},
  pages={101--113},
  year={2022}
}

@inproceedings{thongtanunam2024code,
  title={Code Ownership: The Principles, Differences, and Their Associations with Software Quality},
  author={Thongtanunam, Patanamon and Tantithamthavorn, Chakkrit},
  booktitle={2024 IEEE 35th International Symposium on Software Reliability Engineering (ISSRE)},
  pages={379--390},
  year={2024},
  organization={IEEE}
}

@article{misirli2016studying,
  title={Studying high impact fix-inducing changes},
  author={Misirli, Ayse Tosun and Shihab, Emad and Kamei, Yasukata},
  journal={Empirical Software Engineering},
  volume={21},
  number={2},
  pages={605--641},
  year={2016},
  publisher={Springer}
}

@article{ni2022just,
  title={Just-in-time defect prediction on JavaScript projects: A replication study},
  author={Ni, Chao and Xia, Xin and Lo, David and Yang, Xiaohu and Hassan, Ahmed E},
  journal={ACM Transactions on Software Engineering and Methodology (TOSEM)},
  volume={31},
  number={4},
  pages={1--38},
  year={2022},
  publisher={ACM New York, NY}
}

@inproceedings{khanan2020jitbot,
  title={JITBot: an explainable just-in-time defect prediction bot},
  author={Khanan, Chaiyakarn and Luewichana, Worawit and Pruktharathikoon, Krissakorn and Jiarpakdee, Jirayus and Tantithamthavorn, Chakkrit and Choetkiertikul, Morakot and Ragkhitwetsagul, Chaiyong and Sunetnanta, Thanwadee},
  booktitle={Proceedings of the 35th IEEE/ACM international conference on automated software engineering},
  pages={1336--1339},
  year={2020}
}

@article{fan2019impact,
  title={The impact of mislabeled changes by szz on just-in-time defect prediction},
  author={Fan, Yuanrui and Xia, Xin and Da Costa, Daniel Alencar and Lo, David and Hassan, Ahmed E and Li, Shanping},
  journal={IEEE transactions on software engineering},
  volume={47},
  number={8},
  pages={1559--1586},
  year={2019},
  publisher={IEEE}
}

@inproceedings{li2011case,
  title={A case study of measuring degeneration of software architectures from a defect perspective},
  author={Li, Zude and Long, Jun},
  booktitle={2011 18th Asia-Pacific Software Engineering Conference},
  pages={242--249},
  year={2011},
  organization={IEEE}
}

@inproceedings{bandi2013empirical,
  title={Empirical evidence of code decay: A systematic mapping study},
  author={Bandi, Ajay and Williams, Byron J and Allen, Edward B},
  booktitle={2013 20th Working Conference on Reverse Engineering (WCRE)},
  pages={341--350},
  year={2013},
  organization={IEEE}
}

@article{eick2002does,
  title={Does code decay? assessing the evidence from change management data},
  author={Eick, Stephen G and Graves, Todd L and Karr, Alan F and Marron, J Steve and Mockus, Audris},
  journal={IEEE transactions on software engineering},
  volume={27},
  number={1},
  pages={1--12},
  year={2002},
  publisher={IEEE}
}

@article{shihab2013studying,
  title={Studying re-opened bugs in open source software},
  author={Shihab, Emad and Ihara, Akinori and Kamei, Yasutaka and Ibrahim, Walid M and Ohira, Masao and Adams, Bram and Hassan, Ahmed E and Matsumoto, Ken-ichi},
  journal={Empirical Software Engineering},
  volume={18},
  number={5},
  pages={1005--1042},
  year={2013},
  publisher={Springer}
}

@article{le2013current,
  title={Current challenges in automatic software repair},
  author={Le Goues, Claire and Forrest, Stephanie and Weimer, Westley},
  journal={Software quality journal},
  volume={21},
  number={3},
  pages={421--443},
  year={2013},
  publisher={Springer}
}

@article{hammad2011automatically,
  title={Automatically identifying changes that impact code-to-design traceability during evolution},
  author={Hammad, Maen and Collard, Michael L and Maletic, Jonathan I},
  journal={Software Quality Journal},
  volume={19},
  number={1},
  pages={35--64},
  year={2011},
  publisher={Springer}
}

@inproceedings{jiang2019inferring,
  title={Inferring program transformations from singular examples via big code},
  author={Jiang, Jiajun and Ren, Luyao and Xiong, Yingfei and Zhang, Lingming},
  booktitle={2019 34th IEEE/ACM International Conference on Automated Software Engineering (ASE)},
  pages={255--266},
  year={2019},
  organization={IEEE}
}

@inproceedings{ersoy2016using,
  title={Using hypergraph clustering for software architecture reconstruction of data-tier software},
  author={Ersoy, Ersin and Kaya, Kamer and Alt{\i}n{\i}{\c{s}}{\i}k, Metin and S{\"o}zer, Hasan},
  booktitle={European Conference on Software Architecture},
  pages={326--333},
  year={2016},
  organization={Springer}
}

@article{qiu2024code,
  title={Code multiview hypergraph representation learning for software defect prediction},
  author={Qiu, Shaojian and Huang, Mengyang and Liang, Yun and Peng, Chaoda and Yuan, Yuan},
  journal={IEEE Transactions on Reliability},
  volume={73},
  number={4},
  pages={1863--1876},
  year={2024},
  publisher={IEEE}
}

@inproceedings{wang2024hecs,
  title={HECS: A Hypergraph Learning-Based System for Detecting Extract Class Refactoring Opportunities},
  author={Wang, Luqiao and Wang, Qiangqiang and Wang, Jiaqi and Zhao, Yutong and Wei, Minjie and Quan, Zhou and Cui, Di and Li, Qingshan},
  booktitle={Proceedings of the 33rd ACM SIGSOFT International Symposium on Software Testing and Analysis},
  pages={1851--1855},
  year={2024}
}

@inproceedings{cui2024three,
  title={Three heads are better than one: suggesting move method refactoring opportunities with inter-class code entity dependency enhanced hybrid hypergraph neural network},
  author={Cui, Di and Wang, Jiaqi and Wang, Qiangqiang and Ji, Peng and Qiao, Minglang and Zhao, Yutong and Hu, Jingzhao and Wang, Luqiao and Li, Qingshan},
  booktitle={Proceedings of the 39th IEEE/ACM International Conference on Automated Software Engineering},
  pages={745--757},
  year={2024}
}

@inproceedings{meneely2008predicting,
	title        = {Predicting failures with developer networks and social network analysis},
	author       = {Meneely, Andrew and Williams, Laurie and Snipes, Will and Osborne, Jason},
	year         = 2008,
	booktitle    = {Proceedings of the 16th ACM SIGSOFT International Symposium on Foundations of software engineering},
	pages        = {13--23},
	organization = {ACM}
}

@article{betweenness,
	title        = {A Set of Measures of Centrality Based on Betweenness},
	author       = {Linton C. Freeman},
	year         = 1977,
	journal      = {Sociometry},
	publisher    = {[American Sociological Association, Sage Publications, Inc.]},
	volume       = 40,
	number       = 1,
	pages        = {35--41},
	issn         = {00380431},
	url          = {http://www.jstor.org/stable/3033543}
}

@article{eigenvector,
	title        = {Power and Centrality: A Family of Measures},
	author       = {Bonacich, Phillip},
	year         = 1987,
	journal      = {American Journal of Sociology},
	volume       = 92,
	number       = 5,
	pages        = {1170--1182},
	doi          = {10.1086/228631},
	url          = {https://doi.org/10.1086/228631},
	eprint       = {https://doi.org/10.1086/228631}
}

@article{pedregosa2011scikit,
	title        = {Scikit-learn: Machine learning in Python},
	author       = {Pedregosa, Fabian and Varoquaux, Ga{\"e}l and Gramfort, Alexandre and Michel, Vincent and Thirion, Bertrand and Grisel, Olivier and Blondel, Mathieu and Prettenhofer, Peter and Weiss, Ron and Dubourg, Vincent and others},
	year         = 2011,
	journal      = {the Journal of machine Learning research},
	publisher    = {JMLR. org},
	volume       = 12,
	pages        = {2825--2830}
}

@inproceedings{rong2022modeling,
  title={Modeling review history for reviewer recommendation: A hypergraph approach},
  author={Rong, Guoping and Zhang, Yifan and Yang, Lanxin and Zhang, Fuli and Kuang, Hongyu and Zhang, He},
  booktitle={Proceedings of the 44th international conference on software engineering},
  pages={1381--1392},
  year={2022}
}

@article{wen2018well,
	title        = {How well do change sequences predict defects? sequence learning from software changes},
	author       = {Wen, Ming and Wu, Rongxin and Cheung, Shing-Chi},
	year         = 2018,
	journal      = {IEEE Transactions on Software Engineering},
	publisher    = {IEEE},
	volume       = 46,
	number       = 11,
	pages        = {1155--1175}
}

@inproceedings{kouroshfar2015study,
	title        = {A study on the role of software architecture in the evolution and quality of software},
	author       = {Kouroshfar, Ehsan and Mirakhorli, Mehdi and Bagheri, Hamid and Xiao, Lu and Malek, Sam and Cai, Yuanfang},
	year         = 2015,
	booktitle    = {2015 IEEE/ACM 12th Working Conference on Mining Software Repositories},
	pages        = {246--257},
	organization = {IEEE}
}

@inproceedings{shihab2011high,
	title        = {High-impact defects: a study of breakage and surprise defects},
	author       = {Shihab, Emad and Mockus, Audris and Kamei, Yasutaka and Adams, Bram and Hassan, Ahmed E},
	year         = 2011,
	booktitle    = {Proceedings of the 19th ACM SIGSOFT symposium and the 13th European conference on Foundations of software engineering},
	pages        = {300--310}
}

@article{silva2019co,
	title        = {Co-change patterns: A large scale empirical study},
	author       = {Silva, Luciana L and Valente, Marco Tulio and Maia, Marcelo A},
	year         = 2019,
	journal      = {Journal of Systems and Software},
	publisher    = {Elsevier},
	volume       = 152,
	pages        = {196--214}
}

@inproceedings{d2009relationship,
	title        = {On the relationship between change coupling and software defects},
	author       = {D'Ambros, Marco and Lanza, Michele and Robbes, Romain},
	year         = 2009,
	booktitle    = {2009 16th Working Conference on Reverse Engineering},
	pages        = {135--144},
	organization = {IEEE}
}

@article{zimmermann2005mining,
	title        = {Mining version histories to guide software changes},
	author       = {Zimmermann, Thomas and Zeller, Andreas and Weissgerber, Peter and Diehl, Stephan},
	year         = 2005,
	journal      = {IEEE Transactions on Software Engineering},
	publisher    = {IEEE},
	volume       = 31,
	number       = 6,
	pages        = {429--445}
}

@inproceedings{silva2014assessing,
	title        = {Assessing modularity using co-change clusters},
	author       = {Silva, Luciana Lourdes and Valente, Marco Tulio and Maia, Marcelo de A},
	year         = 2014,
	booktitle    = {Proceedings of the 13th international conference on Modularity},
	pages        = {49--60}
}

@article{silva2015co,
	title        = {Co-change clusters: Extraction and application on assessing software modularity},
	author       = {Silva, Luciana Lourdes and Valente, Marco Tulio and de A. Maia, Marcelo},
	year         = 2015,
	journal      = {Transactions on Aspect-Oriented Software Development XII},
	publisher    = {Springer},
	pages        = {96--131}
}

@article{chawla2002smote,
	title        = {SMOTE: synthetic minority over-sampling technique},
	author       = {Chawla, Nitesh V and Bowyer, Kevin W and Hall, Lawrence O and Kegelmeyer, W Philip},
	year         = 2002,
	journal      = {Journal of artificial intelligence research},
	volume       = 16,
	pages        = {321--357}
}

@inproceedings{kouroshfar2013studying,
	title        = {Studying the effect of co-change dispersion on software quality},
	author       = {Kouroshfar, Ehsan},
	year         = 2013,
	booktitle    = {2013 35th International Conference on Software Engineering (ICSE)},
	pages        = {1450--1452},
	organization = {IEEE}
}

@article{bird2009does,
	title        = {Does distributed development affect software quality? an empirical case study of windows vista},
	author       = {Bird, Christian and Nagappan, Nachiappan and Devanbu, Premkumar and Gall, Harald and Murphy, Brendan},
	year         = 2009,
	journal      = {Communications of the ACM},
	publisher    = {ACM New York, NY, USA},
	volume       = 52,
	number       = 8,
	pages        = {85--93}
}

@inproceedings{yatish2019mining,
	title        = {Mining software defects: Should we consider affected releases?},
	author       = {Yatish, Suraj and Jiarpakdee, Jirayus and Thongtanunam, Patanamon and Tantithamthavorn, Chakkrit},
	year         = 2019,
	booktitle    = {2019 IEEE/ACM 41st international conference on software engineering (ICSE)},
	publisher    = {IEEE Press},
	address      = {Montreal, Quebec, Canada},
	pages        = {654--665},
	organization = {IEEE}
}

@article{gong2021revisiting,
	title        = {Revisiting the impact of dependency network metrics on software defect prediction},
	author       = {Gong, Lina and Rajbahadur, Gopi Krishnan and Hassan, Ahmed E and Jiang, Shujuan},
	year         = 2021,
	journal      = {IEEE Transactions on Software Engineering},
	publisher    = {IEEE},
	volume       = 48,
	number       = 12,
	pages        = {5030--5049}
}

@inproceedings{rahman2013and,
	title        = {How, and why, process metrics are better},
	author       = {Rahman, Foyzur and Devanbu, Premkumar},
	year         = 2013,
	booktitle    = {2013 35th international conference on software engineering (ICSE)},
	publisher    = {IEEE Press},
	address      = {San Francisco, CA, USA},
	pages        = {432--441},
	organization = {IEEE}
}

@article{majumder2022revisiting,
	title        = {Revisiting process versus product metrics: a large scale analysis},
	author       = {Majumder, Suvodeep and Mody, Pranav and Menzies, Tim},
	year         = 2022,
	journal      = {Empirical Software Engineering},
	publisher    = {Springer},
	volume       = 27,
	number       = 3,
	pages        = 60
}

@inproceedings{hassan2009predicting,
	title        = {Predicting faults using the complexity of code changes},
	author       = {Hassan, Ahmed E},
	year         = 2009,
	booktitle    = {2009 IEEE 31st international conference on software engineering},
	publisher    = {IEEE Computer Society},
	address      = {USA},
	pages        = {78--88},
	organization = {IEEE}
}

@article{majumder2024less,
	title        = {When less is more: on the value of “co-training” for semi-supervised software defect predictors},
	author       = {Majumder, Suvodeep and Chakraborty, Joymallya and Menzies, Tim},
	year         = 2024,
	journal      = {Empirical Software Engineering},
	publisher    = {Springer},
	volume       = 29,
	number       = 2,
	pages        = 51
}

@inproceedings{nagappan2010change,
	title        = {Change bursts as defect predictors},
	author       = {Nagappan, Nachiappan and Zeller, Andreas and Zimmermann, Thomas and Herzig, Kim and Murphy, Brendan},
	year         = 2010,
	booktitle    = {2010 IEEE 21st international symposium on software reliability engineering},
	publisher    = {IEEE Computer Society},
	address      = {USA},
	pages        = {309--318},
	organization = {IEEE}
}

@article{wang2013using,
	title        = {Using class imbalance learning for software defect prediction},
	author       = {Wang, Shuo and Yao, Xin},
	year         = 2013,
	journal      = {IEEE Transactions on Reliability},
	publisher    = {IEEE},
	volume       = 62,
	number       = 2,
	pages        = {434--443}
}

@inproceedings{pinzger2008can,
	title        = {Can developer-module networks predict failures?},
	author       = {Pinzger, Martin and Nagappan, Nachiappan and Murphy, Brendan},
	year         = 2008,
	booktitle    = {Proceedings of the 16th ACM SIGSOFT International Symposium on Foundations of software engineering},
	pages        = {2--12}
}

@article{kamei2012large,
	title        = {A large-scale empirical study of just-in-time quality assurance},
	author       = {Kamei, Yasutaka and Shihab, Emad and Adams, Bram and Hassan, Ahmed E and Mockus, Audris and Sinha, Anand and Ubayashi, Naoyasu},
	year         = 2012,
	journal      = {IEEE Transactions on Software Engineering},
	publisher    = {IEEE},
	volume       = 39,
	number       = 6,
	pages        = {757--773}
}

@inproceedings{tomani2023beyond,
	title        = {Beyond in-domain scenarios: robust density-aware calibration},
	author       = {Tomani, Christian and Waseda, Futa Kai and Shen, Yuesong and Cremers, Daniel},
	year         = 2023,
	booktitle    = {International Conference on Machine Learning},
	pages        = {34344--34368},
	organization = {PMLR}
}

@article{singhalfoundation,
  title   = {Foundation Vision Models are Unsupervised Image Canonicalizers},
  author  = {Singhal, Utkarsh and Feng, Ryan and Yu, Stella X. and Prakash, Atul}
}

@article{gong2022comprehensive,
	title        = {A comprehensive investigation of the impact of class overlap on software defect prediction},
	author       = {Gong, Lina and Zhang, Haoxiang and Zhang, Jingxuan and Wei, Mingqiang and Huang, Zhiqiu},
	year         = 2022,
	journal      = {IEEE transactions on software engineering},
	publisher    = {IEEE},
	volume       = 49,
	number       = 4,
	pages        = {2440--2458}
}

@article{shin2021explainable,
	title        = {Explainable software defect prediction: Are we there yet?},
	author       = {Shin, Jiho and Aleithan, Reem and Nam, Jaechang and Wang, Junjie and Wang, Song},
	year         = 2021,
	journal      = {arXiv preprint arXiv:2111.10901}
}

@inproceedings{rajbahadur2017impact,
	title        = {The impact of using regression models to build defect classifiers},
	author       = {Rajbahadur, Gopi Krishnan and Wang, Shaowei and Kamei, Yasutaka and Hassan, Ahmed E},
	year         = 2017,
	booktitle    = {2017 IEEE/ACM 14th International Conference on Mining Software Repositories (MSR)},
	pages        = {135--145},
	organization = {IEEE}
}

@article{tantithamthavorn2016empirical,
	title        = {An empirical comparison of model validation techniques for defect prediction models},
	author       = {Tantithamthavorn, Chakkrit and McIntosh, Shane and Hassan, Ahmed E and Matsumoto, Kenichi},
	year         = 2016,
	journal      = {IEEE Transactions on Software Engineering},
	publisher    = {IEEE},
	volume       = 43,
	number       = 1,
	pages        = {1--18}
}

@article{wang2022machine,
	title        = {Machine/deep learning for software engineering: A systematic literature review},
	author       = {Wang, Simin and Huang, Liguo and Gao, Amiao and Ge, Jidong and Zhang, Tengfei and Feng, Haitao and Satyarth, Ishna and Li, Ming and Zhang, He and Ng, Vincent},
	year         = 2022,
	journal      = {IEEE Transactions on Software Engineering},
	publisher    = {IEEE},
	volume       = 49,
	number       = 3,
	pages        = {1188--1231}
}

@article{lee2025survey,
	title        = {A survey on hypergraph mining: Patterns, tools, and generators},
	author       = {Lee, Geon and Bu, Fanchen and Eliassi-Rad, Tina and Shin, Kijung},
	year         = 2025,
	journal      = {ACM Computing Surveys},
	publisher    = {ACM New York, NY},
	volume       = 57,
	number       = 8,
	pages        = {1--36}
}

@inproceedings{pan2023fine,
	title        = {Fine-grained commit-level vulnerability type prediction by CWE tree structure},
	author       = {Pan, Shengyi and Bao, Lingfeng and Xia, Xin and Lo, David and Li, Shanping},
	year         = 2023,
	booktitle    = {2023 IEEE/ACM 45th International Conference on Software Engineering (ICSE)},
	pages        = {957--969},
	organization = {IEEE}
}

@inproceedings{ni2022best,
	title        = {The best of both worlds: integrating semantic features with expert features for defect prediction and localization},
	author       = {Ni, Chao and Wang, Wei and Yang, Kaiwen and Xia, Xin and Liu, Kui and Lo, David},
	year         = 2022,
	booktitle    = {Proceedings of the 30th ACM Joint European Software Engineering Conference and Symposium on the Foundations of Software Engineering},
	pages        = {672--683}
}

@inproceedings{keshavarz2022apachejit,
	title        = {Apachejit: a large dataset for just-in-time defect prediction},
	author       = {Keshavarz, Hossein and Nagappan, Meiyappan},
	year         = 2022,
	booktitle    = {Proceedings of the 19th international conference on mining software repositories},
	pages        = {191--195}
}

@inproceedings{mcintosh2018fix,
	title        = {Are fix-inducing changes a moving target? a longitudinal case study of just-in-time defect prediction},
	author       = {McIntosh, Shane and Kamei, Yasutaka},
	year         = 2018,
	booktitle    = {Proceedings of the 40th international conference on software engineering},
	pages        = {560--560}
}

@inproceedings{mahbub2023defectors,
	title        = {Defectors: A large, diverse python dataset for defect prediction},
	author       = {Mahbub, Parvez and Shuvo, Ohiduzzaman and Rahman, Mohammad Masudur},
	year         = 2023,
	booktitle    = {2023 IEEE/ACM 20th International Conference on Mining Software Repositories (MSR)},
	pages        = {393--397},
	organization = {IEEE}
}

@article{zhou2025bridging,
	title        = {Bridging expert knowledge with deep learning techniques for just-in-time defect prediction},
	author       = {Zhou, Xin and Han, DongGyun and Lo, David},
	year         = 2025,
	journal      = {Empirical Software Engineering},
	publisher    = {Springer},
	volume       = 30,
	number       = 1,
	pages        = {1--44}
}

@article{sliwerski2005changes,
	title        = {When do changes induce fixes?},
	author       = {{\'S}liwerski, Jacek and Zimmermann, Thomas and Zeller, Andreas},
	year         = 2005,
	journal      = {ACM sigsoft software engineering notes},
	publisher    = {ACM New York, NY, USA},
	volume       = 30,
	number       = 4,
	pages        = {1--5}
}

@article{kirbas2017relationship,
	title        = {The relationship between evolutionary coupling and defects in large industrial software},
	author       = {Kirbas, Serkan and Caglayan, Bora and Hall, Tracy and Counsell, Steve and Bowes, David and Sen, Alper and Bener, Ayse},
	year         = 2017,
	journal      = {Journal of Software: Evolution and Process},
	publisher    = {Wiley Online Library},
	volume       = 29,
	number       = 4,
	pages        = {e1842}
}

@article{hrishikesh2025co,
	title        = {Co-Change Graph Entropy: A New Process Metric for Defect Prediction},
	author       = {Hrishikesh, Ethari and Kumar, Amit and Bhardwaj, Meher and Agarwal, Sonali},
	year         = 2025,
	journal      = {arXiv preprint arXiv:2504.18511}
}

@inproceedings{nguyen2010studying,
	title        = {Studying the impact of dependency network measures on software quality},
	author       = {Nguyen, Thanh HD and Adams, Bram and Hassan, Ahmed E},
	year         = 2010,
	booktitle    = {2010 IEEE International Conference on Software Maintenance},
	pages        = {1--10},
	organization = {IEEE}
}

@article{amasaki2020cross,
	title        = {Cross-version defect prediction: use historical data, cross-project data, or both?},
	author       = {Amasaki, Sousuke},
	year         = 2020,
	journal      = {Empirical Software Engineering},
	publisher    = {Springer},
	volume       = 25,
	pages        = {1573--1595}
}

@article{chidamber1994metrics,
	title        = {A metrics suite for object oriented design},
	author       = {Chidamber, Shyam R and Kemerer, Chris F},
	year         = 1994,
	journal      = {IEEE Transactions on software engineering},
	publisher    = {IEEE},
	volume       = 20,
	number       = 6,
	pages        = {476--493}
}

@article{mccabe1976complexity,
	title        = {A complexity measure},
	author       = {McCabe, Thomas J},
	year         = 1976,
	journal      = {IEEE Transactions on software Engineering},
	publisher    = {IEEE},
	number       = 4,
	pages        = {308--320}
}

@inproceedings{moser2008comparative,
	title        = {A comparative analysis of the efficiency of change metrics and static code attributes for defect prediction},
	author       = {Moser, Raimund and Pedrycz, Witold and Succi, Giancarlo},
	year         = 2008,
	booktitle    = {Proceedings of the 30th international conference on Software engineering},
	pages        = {181--190}
}

@inproceedings{zimmermann2008predicting,
	title        = {Predicting defects using network analysis on dependency graphs},
	author       = {Zimmermann, Thomas and Nagappan, Nachiappan},
	year         = 2008,
	booktitle    = {Proceedings of the 30th international conference on Software engineering},
	pages        = {531--540}
}

@inproceedings{yao2020assessing,
	title        = {Assessing software defection prediction performance: Why using the Matthews correlation coefficient matters},
	author       = {Yao, Jingxiu and Shepperd, Martin},
	year         = 2020,
	booktitle    = {Proceedings of the 24th International Conference on Evaluation and Assessment in Software Engineering},
	pages        = {120--129}
}

@article{yamada2014high,
	title        = {High-dimensional feature selection by feature-wise kernelized lasso},
	author       = {Yamada, Makoto and Jitkrittum, Wittawat and Sigal, Leonid and Xing, Eric P and Sugiyama, Masashi},
	year         = 2014,
	journal      = {Neural computation},
	publisher    = {MIT Press},
	volume       = 26,
	number       = 1,
	pages        = {185--207}
}

@inproceedings{xu2016impact,
	title        = {The impact of feature selection on defect prediction performance: An empirical comparison},
	author       = {Xu, Zhou and Liu, Jin and Yang, Zijiang and An, Gege and Jia, Xiangyang},
	year         = 2016,
	booktitle    = {2016 IEEE 27th international symposium on software reliability engineering (ISSRE)},
	pages        = {309--320},
	organization = {IEEE}
}

@article{kovalenko2022vector,
  title={Vector centrality in hypergraphs},
  author={Kovalenko, Kirill and Romance, Miguel and Vasilyeva, Ekaterina and Aleja, David and Criado, Regino and Musatov, Daniil and Raigorodskii, Andrei M and Flores, Julio and Samoylenko, Ivan and Alfaro-Bittner, Karin and others},
  journal={Chaos, Solitons \& Fractals},
  volume={162},
  pages={112397},
  year={2022},
  publisher={Elsevier}
}

\end{document}